\documentclass[preprintnumbers,amsmath,amsart,amssymb,floatfix,11pt,prd,onecolumn,
superscriptaddress,nofootinbib]{revtex4}
\usepackage{latexsym}
\usepackage{epsfig}
\usepackage{epstopdf}
\usepackage{amssymb}
\usepackage{amsmath, hyperref}
\usepackage{epsf,graphics,graphicx}
\usepackage{float}

\begin{document}

\title{The Effects of Running Gravitational Coupling On Rotating Black Holes}

\author{Sumarna Haroon}\email{sumarna.haroon@sns.nust.edu.pk}
\affiliation{School of Natural Sciences, National University of
Sciences and Technology, Islamabad, 44000, Pakistan}
\author{Mubasher Jamil}\email{mjamil@sns.nust.edu.pk}
\affiliation{School of Natural Sciences, National University of
Sciences and Technology, Islamabad, 44000, Pakistan}
\author{Kai Lin}
\affiliation{Institute for Advanced Physics and Mathematics,
Zhejiang University of Technology, Hangzhou 310032, China}
\author{Petar Pavlovic}\email{petar.pavlovic@desy.de}
\affiliation{Institut f\"{u}r Theoretische Physik, Universit\"{a}t
Hamburg, Luruper Chaussee 149, 22761 Hamburg, Germany}
\author{Marko Sossich}\email{marko.sossich@fer.hr}
\affiliation{University of Zagreb, Faculty of Electrical Engineering and Computing, Department of Applied Physics, Unska 3, 10 000 Zagreb, Croatia}
\author{Anzhong Wang}
\affiliation{Institute for Advanced Physics and Mathematics,
Zhejiang University of Technology, Hangzhou 310032, China}
\affiliation{GCAP-CASPER, Physics Department, Baylor University,
Waco, TX 76798-7316, USA}

\begin{abstract}
{\bf Abstract:} In this work we investigate the consequences of running gravitational coupling on the properties of rotating black holes. 
Apart from the changes induced in the space-time structure of such black holes, we also study the  implications to Penrose process and geodetic precession. We are motivated by the functional form of gravitational coupling previously investigated in the context of infra-red limit 
of asymptotic safe gravity theory. In this approach, the involvement of a new parameter $\tilde{\xi}$ in this solution makes it different from Schwarzschild black hole. 
 The Killing horizon, event horizon and singularity of the computed metric is then discussed. It is noticed that the ergosphere is increased as $\tilde{\xi}$ increases. Considering the black hole solution in equatorial plane, the geodesics of particles, both null and time like cases, are explored. The effective potential is computed and graphically analyzed for different values of parameter $\tilde{\xi}$. The energy extraction from black hole is investigated via Penrose process. For the same values of spin parameter, the numerical results suggest that the efficiency of Penrose process is greater in quantum corrected gravity than in Kerr Black Hole. At the end, a brief discussion on Lense-Thirring frequency is also done.
\end{abstract}

\maketitle

\newpage

\section{Introduction}

The problem of finding a consistent theory of quantum gravity remains to be the central challenge in theoretical physics. During the past
decades different approaches and perspectives on this issue were developed, such as loop quantum gravity \cite{loop1, loop2}, string theory \cite{string1, string2},
and effective approaches of modified gravity theories \cite{mod1, mod2, mod3}. These attempts also addressed different more specific problems of cosmology and astrophysics
-- including dark energy and dark matter problem, the horizon problem, as well as the black hole physics and singularities of general relativity.
All of these problems are connected to the potential limitations of Einstein's general relativity, and are therefore important motivation
and reference in the investigation of quantum gravity.\\
It is usually assumed that Einstein's general theory of relativity is valid only as an effective theory of gravity. According to this
picture, general relativity can be taken as a correct description of gravitational interaction only up to certain scales of energy and characteristic
distances. When they get comparable to Planck scale the theory is expected to break, and to be replaced with a completely different physical model.
This reasoning seems to be supported by the well known fact that the Einstein-Hilbert action, leading to the field equations of general
relativity, is perturbative non-renormalizable \cite{bern}. However, Weinberg proposed a new nonperturbative notion of renormalizibility
which is called ``asymptotic safety'' \cite{wein}, based on the existence of a nontrivial fixed point in renormalization group, which makes the physical couplings of the theory non-divergent. The basic assumption of Weinberg's proposal was that gravity can meet this criteria,
and thus its description can be considered as a consistent field theory on all scales. A review and discussion of attempts to prove the existence of
this fixed point for gravity can be found in \cite{nied}.\\
In the perspective of research on quantum gravity, it is of special interest to consider the consequences of the asymptotically safe
gravity picture on the well known physical systems, which are in principle also accessible to observations. Black holes are a good example
of such system, where the corrections to standard description of gravity could be important. Black holes in asymptotically safe gravity
were previously studied in \cite{br, br3, koch1, koch2, P, bhole1, bhole2, DF1, DF2, DF3}. In this work we continue the investigation
of black holes in asymptotically safe gravity, considering the rotating black hole solutions, and focusing on the functional form of gravitational 
coupling inspired by the potential infra-red limit of the theory,
due to its observational relevance. Previous to this work the quantum gravity effects in the Kerr spacetime were studied in \cite{RT},
where the structure of horizons, the ergosphere, the Penrose process and the static limit surfaces were investigated considering the
generalization of gravitational constant to a general function of radial coordinate, $G(r)$ -- that comes as a result of quantum effects.
Our current work could be understood as a further extension of the analysis performed in \cite{RT}. While focusing on a more specific setting
of infra-red limit of asymptotic safe gravity, we concentrate on a specific form of $G(r)$ function, which now enables us to obtain the concrete
solutions for equatorial geodesics, Penrose process, and to analyze the Lense-Thiring effect. \\ \\

The study of geodesics for both null (photon) and time-like (massive) particles, has always held a significant importance. The analysis of circular motion of particles in a curved space time exhibits its geometrical behaviour. For some recent work on geodesics see \cite{BB}. For better understanding of the technique used in this paper for the computation of geodesics of particles, see \cite{C}. The circular orbits and the associated orbits are then discussed in detail. The angular velocity and effective potential are computed and graphically demonstrated. \\ \\
Energy is a conserved quantity on the spacetime of stationary rotating black hole, due to the existence of associated Killing vector, $K_{\mu}=\partial_{t}$, so that
$E= - K^{\mu}u^{\mu}$, where $u^{\mu}$ is a four-velocity defined on some geodesic. At the asymptotic infinity both $K^{\mu}$ and $u^{\mu}$ are timelike, so that energy is always positive.
However, Killing vector becomes null at the Killing horizon, and spacelike inside the region known as ergosphere -- which represents the space between the outer event horizon
and the Killing horizon of a black hole. It is therefore possible that energy becomes  negative quantity in the
ergosphere of a stationary rotating black hole. This fact was used by Penrose, who proposed a mechanism of extraction of energy from Kerr black hole. Starting from a particle
falling into black hole -- which is of course defined by the positive energy -- one can consider the case where it decays in the ergosphere, into one particle carrying positive energy, and the other
particle carrying negative energy.
Since the total energy needs to be conserved, if we assume that the negative energy particle crosses the event horizon and the positive energy particle leaves the ergosphere
reaching the observer, its energy will be higher than the energy of the initial particle (since it is the difference of the initial energy and the negative energy of the second particle).
In this work we will analyze the Penrose process considering the running gravitational coupling, and also discuss the consequences on the maximal efficiency of Penrose process. \\

A rotational sphere with electric charge generates magnetic field. Similarly, it is expected that a rotating black hole, or a massive star, also produces ``magnetic effect" of gravity according to modern gravitational theory. Such phenomenon is known as Lense-Thirring effect which was firstly proposed by Lense and Thirring in 1918 \cite{LenseThirring}. It can be shown that this effect is beyond Newtonian gravity. In this work, we also investigate the Lense-Thirring effect for the rotating black hole considering the varying Newtonian coupling. Our study found that the relevant properties are determined by $a$, $\theta$ and $M$, and in particular, by the new parameter $\tilde{\xi}$.\\

The outline of this paper is established as follows. In section II, some already discussed concepts are being outlined briefly. In section III, a black hole solution in IR regime with asymptotically safe gravity theory is constructed, following with the comments on event horizon and singularity of the computed rotating metric. In section IV, equatorial null and time-like geodesics of this black hole are taken into account along with the discussion on effective potential. Section IV is on the study of Penrose process. Section V gives a thorough description on Lense-Thirring effect, and we summarize our findings in section VI.
\section{A Review}
This section is actually a brief discussion on running Newton coupling which arise in the framework of asymptotically safe gravity theory. Our main interest is in its infra red (IR) limit. A quantum corrected Kerr metric is also recalled. The key point to mention here is that when quantum corrections are applied to black hole spacetime, the system is modified, such that the Newton constant $G$ turns into a $r$-dependent running Newton coupling $G(r)$ i.e.
\begin{eqnarray}\nonumber
G\rightarrow{G(r)}.
\end{eqnarray}
\subsection{The Running Coupling in Asymptotically Safe Gravity}
 The solution of the renormalization group (RG) equation for the running Newton coupling, $G(p)$, of the action (\ref{action}) is computed in \cite{RG}, using 
 the one-loop correction:
\begin{eqnarray}
G(p)=\frac{G_N}{1+\xi{p^2}G_N},
\end{eqnarray}
where $G_N$ is Newton's constant at classical level. For simplicity, the value of $G_N$ will be equated to unity in rest of our analysis. Here $\xi$ is also a coupling coefficient.\\
Cai and Easson broadened the study of black hole solution in safe gravity by considering higher derivative terms in their analysis \cite{P}. They initiated their study by introducing an effective action
\begin{eqnarray}\nonumber
\Gamma_p[g_{\mu\nu}]&=&\int{d^4}x\sqrt{-g}\Bigg[p^4g_o(p)+p^2g_1(p){\cal{R}}+g_{2a}(p){\cal{R}}^2\\\label{action}
&+&g_{2b}(p){\cal{R}}_{\mu\nu}{\cal{R}}^{\mu\nu}+g_{2c}(p){\cal{R}}_{\mu\nu\sigma\rho}{\cal{R}}^{\mu\nu\sigma\rho}+{\cal{O}}(p^{-2}{\cal{R}}^3)+...\Bigg],
\end{eqnarray}
where $g_{\mu\nu}$ represents metric tensor with $g$ as its determinant, Ricci scalar, Ricci tensor and Riemann tensor are denoted by ${\cal{R}}$, ${\cal{R}}_{\mu\nu}$ and ${\cal{R}}_{\mu\nu\sigma\rho}$, respectively,
 $p$ is the momentum cutoff and $g_i~(0,1,2a,...)$ are dimensionless running couplings satisfying the renormalization group (RG) equations, for example:
 \begin{equation}
 g_{0}(p)=-\frac{\Lambda(p)}{8 \pi G_{N}(p)}p^{-4}, \qquad g_{1}(p)=\frac{1}{8 \pi G_{N}(p)}p^{-2}, \qquad \frac{d}{d \ln p}g_{i}(p)=\beta_{i}(g).
 \end{equation}

Further, it was shown that for large values of radial coordinate $r$, the momentum cut-off goes asymptotically small i.e. $p\sim 1/r$; it may go below the Planck scale. Under this limit (so-called Infra-red or IR), the running Newton coupling $G(r)$ takes the form
\begin{equation}\label{gtt}
G(r)\simeq\left(1-\frac{\tilde\xi}{r^2}\right),
\end{equation}
for $r \gg l_{Planck}$, where $\tilde{\xi}$ differs from $\xi$ by $\mathcal{O}(1)$ and has constant value, less than unity. For more understanding of this running coupling, it is recommended to see \cite{P}. 
Although the main motivation of this form of $G(r)$ comes from the potential IR limit of asymptotic safe gravity, we note that the analysis 
performed in this work can be understood as more general and not only limited to the asymptotic safety program, since the basic 
assumption of its validity is only that the quantum effects can be described by the correction given in (\ref{gtt}). 
\subsection{RG Improved Kerr Metric}
The analysis of RG improved Schwarzschild metric is done in \cite{BR}, and it was noticed that apart from usual Schwarzschild horizon, the presence of a new horizon was noticed which, at critical mass, coincides with the outer horizon. To understand the technique used for the computation of RG improved Schwarzschild metric it is suggested to see \cite{BR}. With the help of similar analysis an improved Kerr metric was suggested by Reuter and Tuiran \cite{RT}. In their analysis they considered Newton's coupling $G$ to be $r$-dependent i.e. $G=G(r)$. With this assumption they arrived at the improved Kerr metric form as
\begin{eqnarray}\label{kerr1}
ds^2&=&-\left(1-\frac{2Mr}{\Sigma}G(r)\right)dt^2-\frac{4aMr\sin^2\theta}{\Sigma}G(r)dtd\phi+\frac{\Sigma}{\Delta}dr^2+\Sigma{d\theta^2}\\\nonumber
&+&\sin^2\theta\Bigg[r^2+a^2+\frac{2a^2Mr\sin^2\theta}{\Sigma}G(r)\Bigg]d\phi^2,
\end{eqnarray}
where $\Delta=r^2-2MrG(r)+a^2$ and $\Sigma=r^2+a^2\cos^2\theta$. Here $a$ is defined as rotational parameter.\\
\section{Kerr Metric in the Infra-red limit of Asymptotically Safe Gravity Theory }
Considering the running Newton parameter (\ref{gtt}) in above metric, we reach to the form of improved Kerr metric solution in asymptotically safe gravity in the infra-red limit. The metric is thus given by
\begin{eqnarray}\nonumber
ds^2&=&-\Big(1-\frac{2Mr}{\Sigma}\Big(1-\frac{\tilde{\xi}}{r^2}\Big)\Big)dt^2-\frac{4aMr\sin^2\theta}{\Sigma}\Big(1-\frac{\tilde{\xi}}{r^2}\Big)dtd\phi+\frac{\Sigma}{\Delta}dr^2+\Sigma{d\theta^2}\\\label{metric}
&+&\sin^2\theta\bigg[r^2+a^2+\frac{2a^2Mr}{\Sigma}\sin^2\theta\Big(1-\frac{\tilde{\xi}}{r^2}\Big)\bigg]d\phi^2,
\end{eqnarray}
where $\Delta=r^2-2Mr+\frac{2M\tilde{\xi}}{r}+a^2$.
This metric reduces to its static and spherically symmetric version when $a\rightarrow{0}$. For reader's better understanding and to provide stronger grounds for the results computed in rest of the sections, a detail derivation of metric (\ref{metric}), using the technique in \cite{A}, is presented in appendix. In next sections we are going to take into account some other characteristic behaviours of metric (\ref{metric}).
\subsection{Event and Killing horizons}
\begin{figure}
    \centering
    \includegraphics[width=0.46\textwidth]{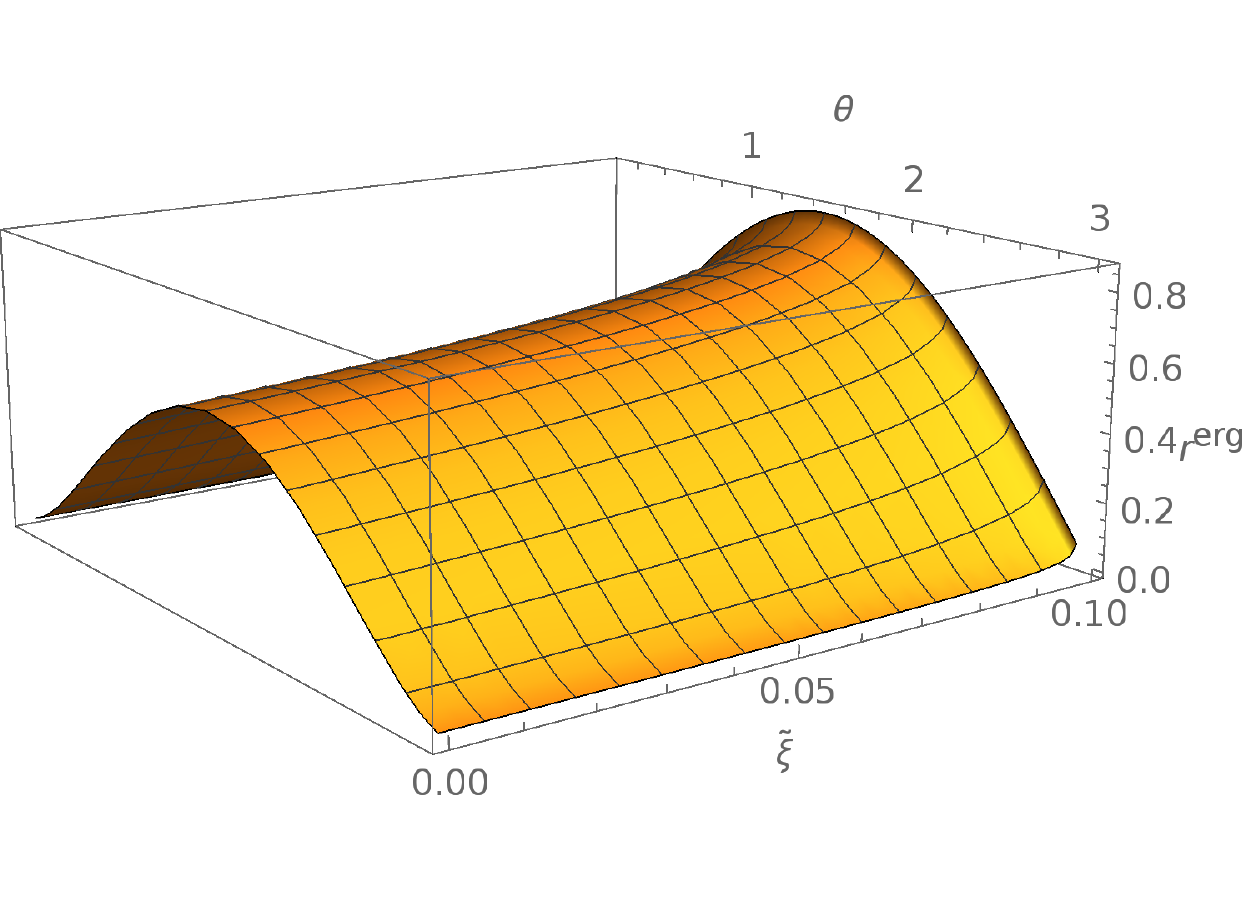}
    \caption{\scriptsize{Difference between the Killing horizon and outer event horizon, $r^{erg}$, in the IR limit of quantum corrected gravitational coupling,
    for the black hole defined by $a=0.9$ and $M=1$.
   }}\label{Fig1}
\end{figure}

\begin{figure}
    \centering
    \includegraphics[width=0.46\textwidth]{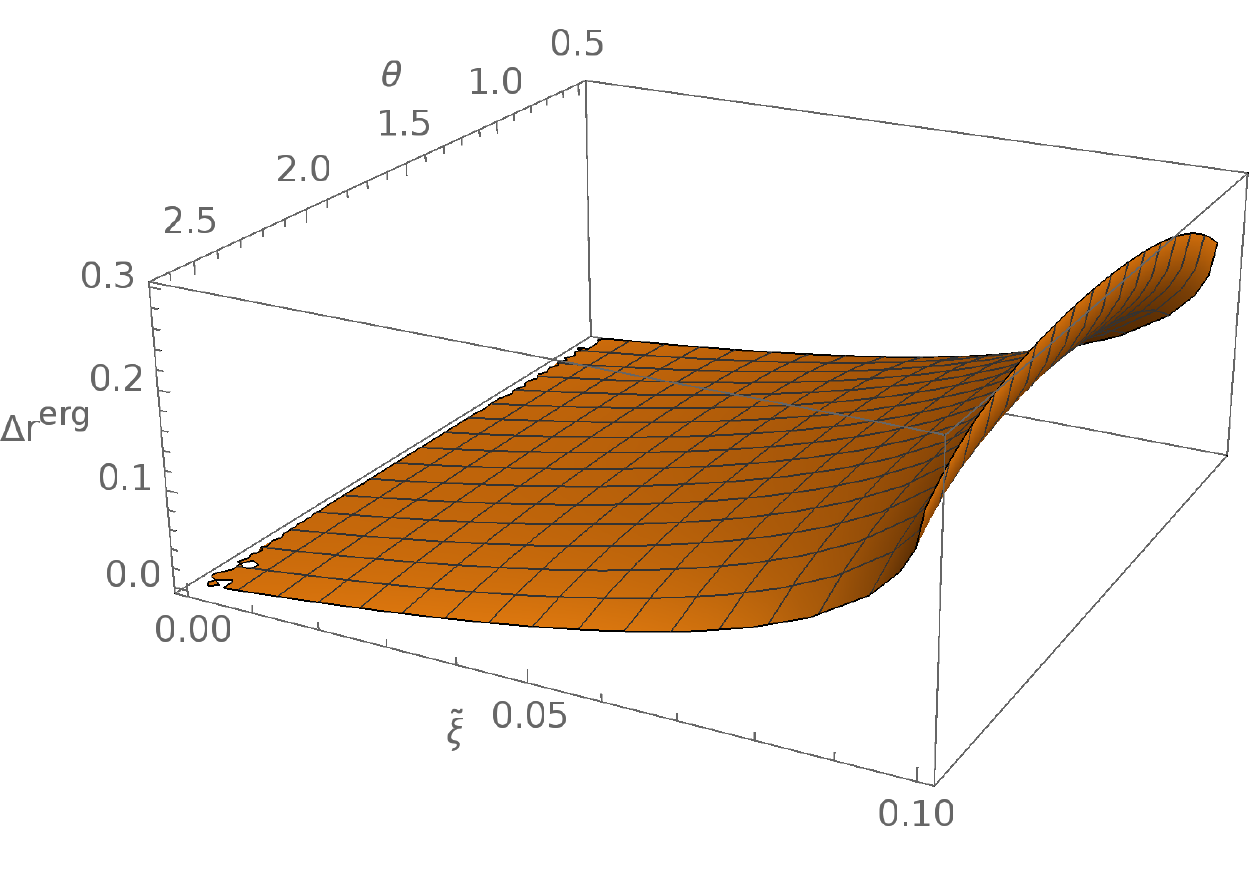}
    \caption{\scriptsize{Difference between $r^{erg}$ in the IR limit of quantum corrected gravitational coupling and GR which we label as $\Delta r^{erg}$. The black hole is defined by $a=0.9$ and $M=1$.
   }}\label{Fig2}
\end{figure}

\begin{figure}[H]
    \includegraphics[width=0.22\linewidth]{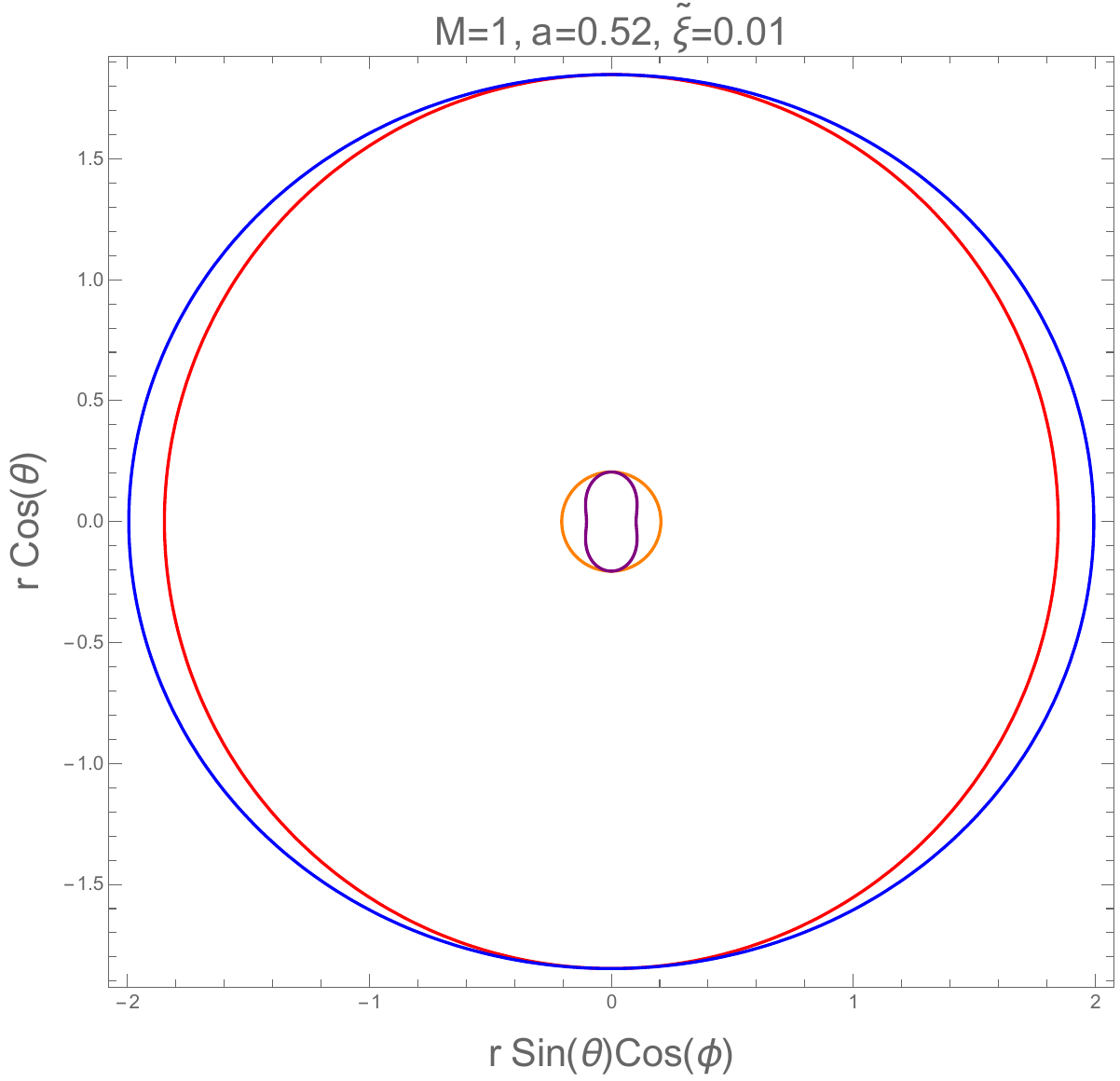}
    \includegraphics[width=0.22\linewidth]{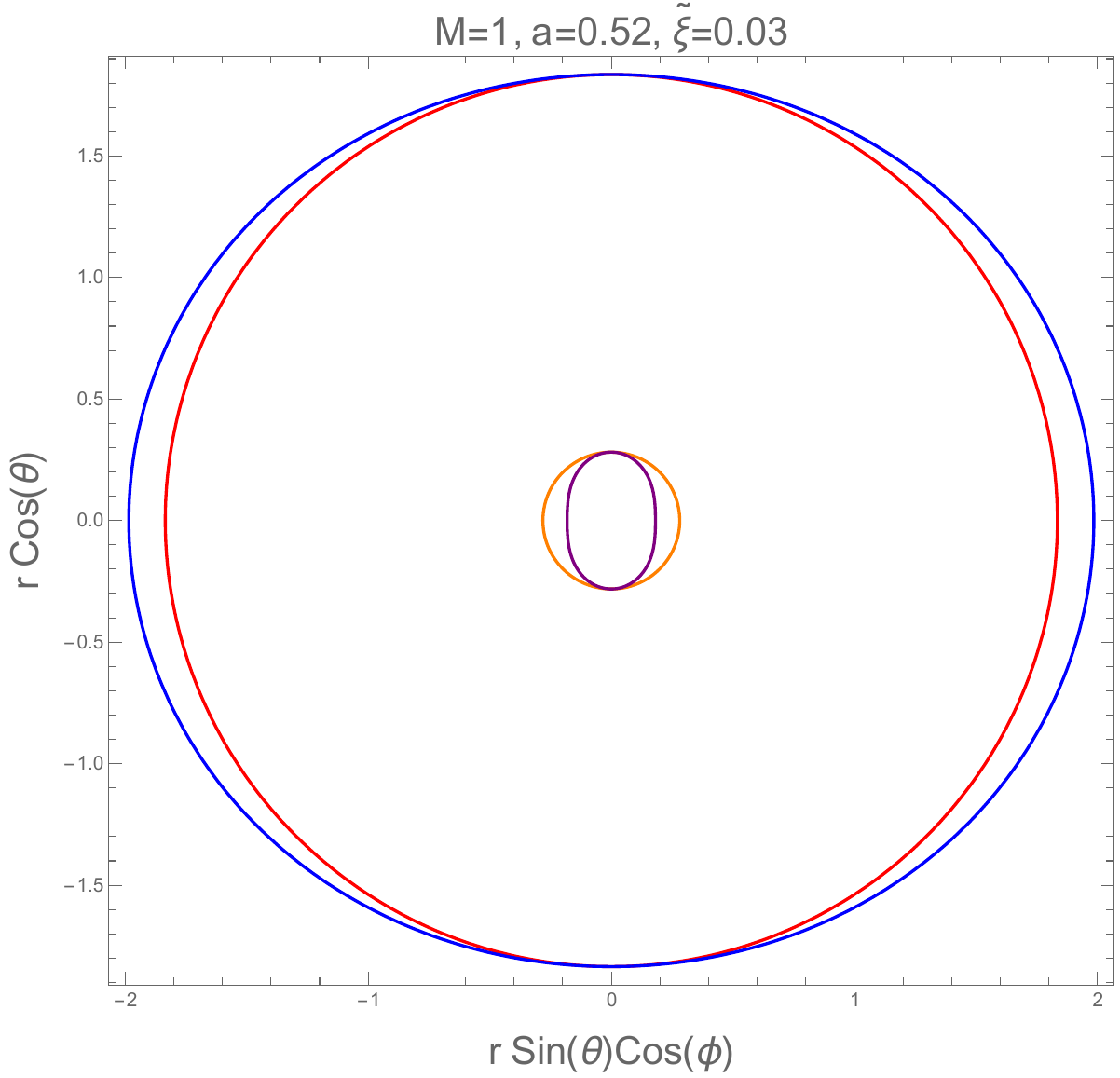}
    \includegraphics[width=0.22\linewidth]{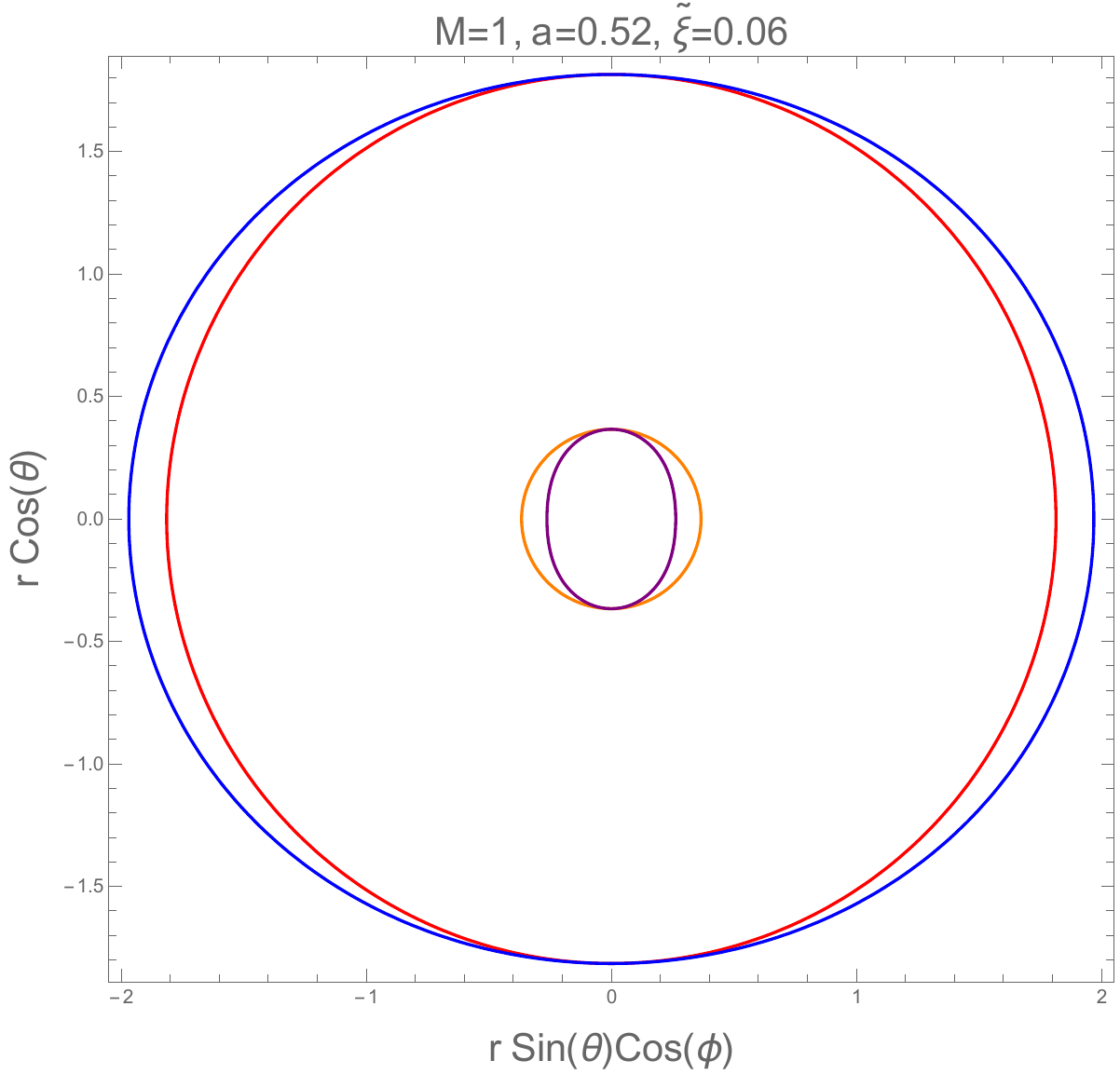}
    \includegraphics[width=0.22\linewidth]{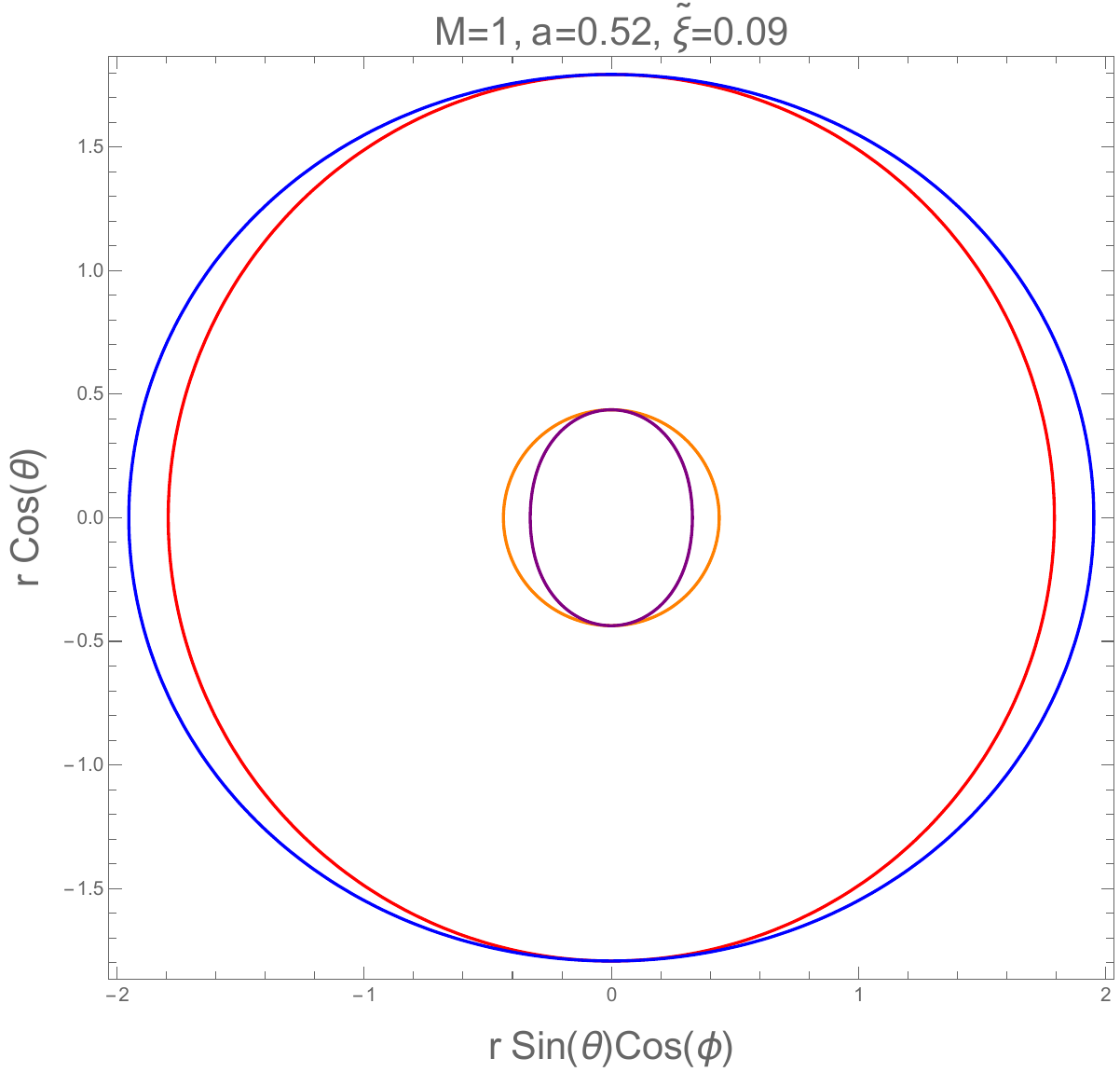}
     \includegraphics[width=0.22\linewidth]{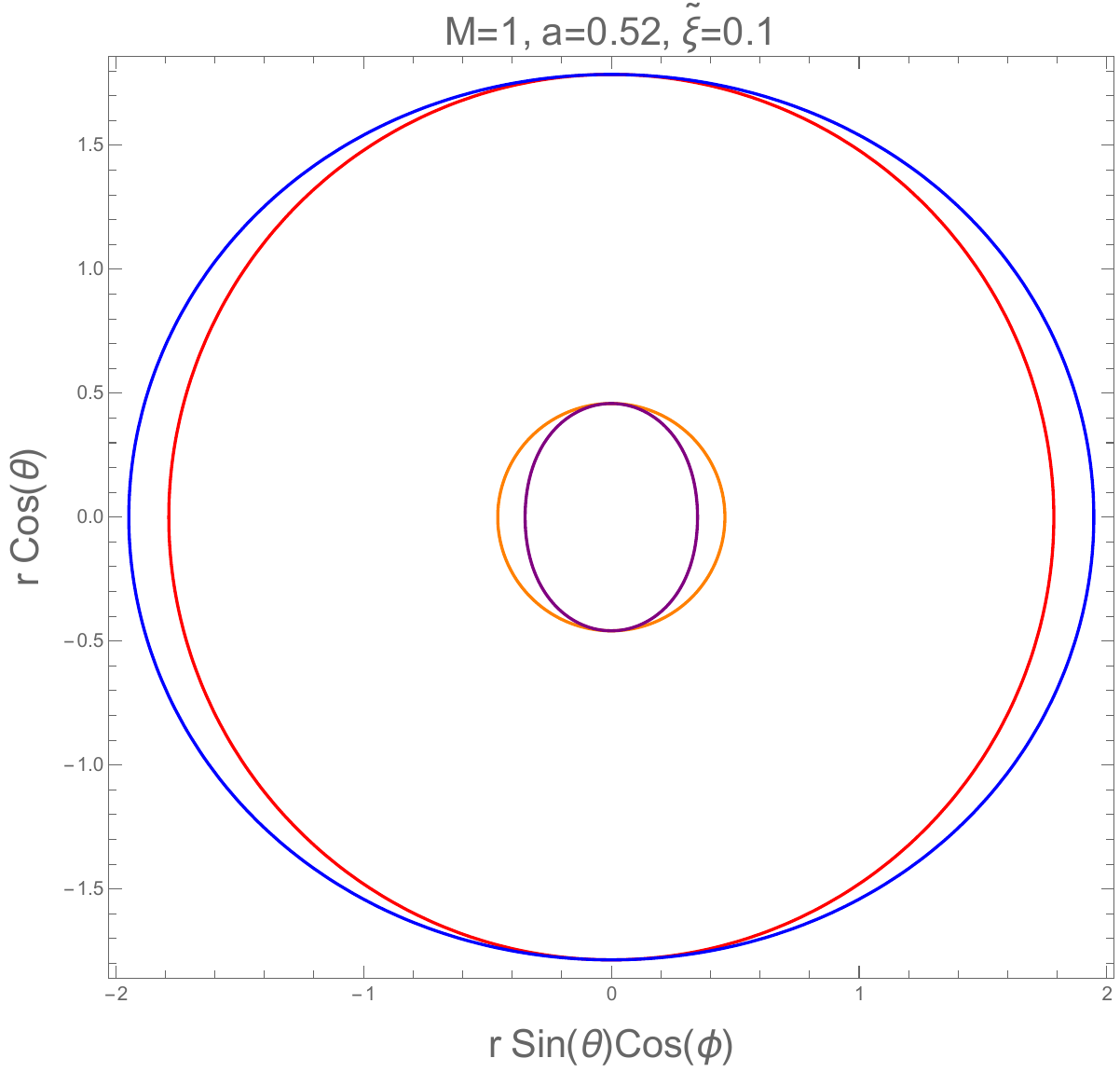}
      \includegraphics[width=0.22\linewidth]{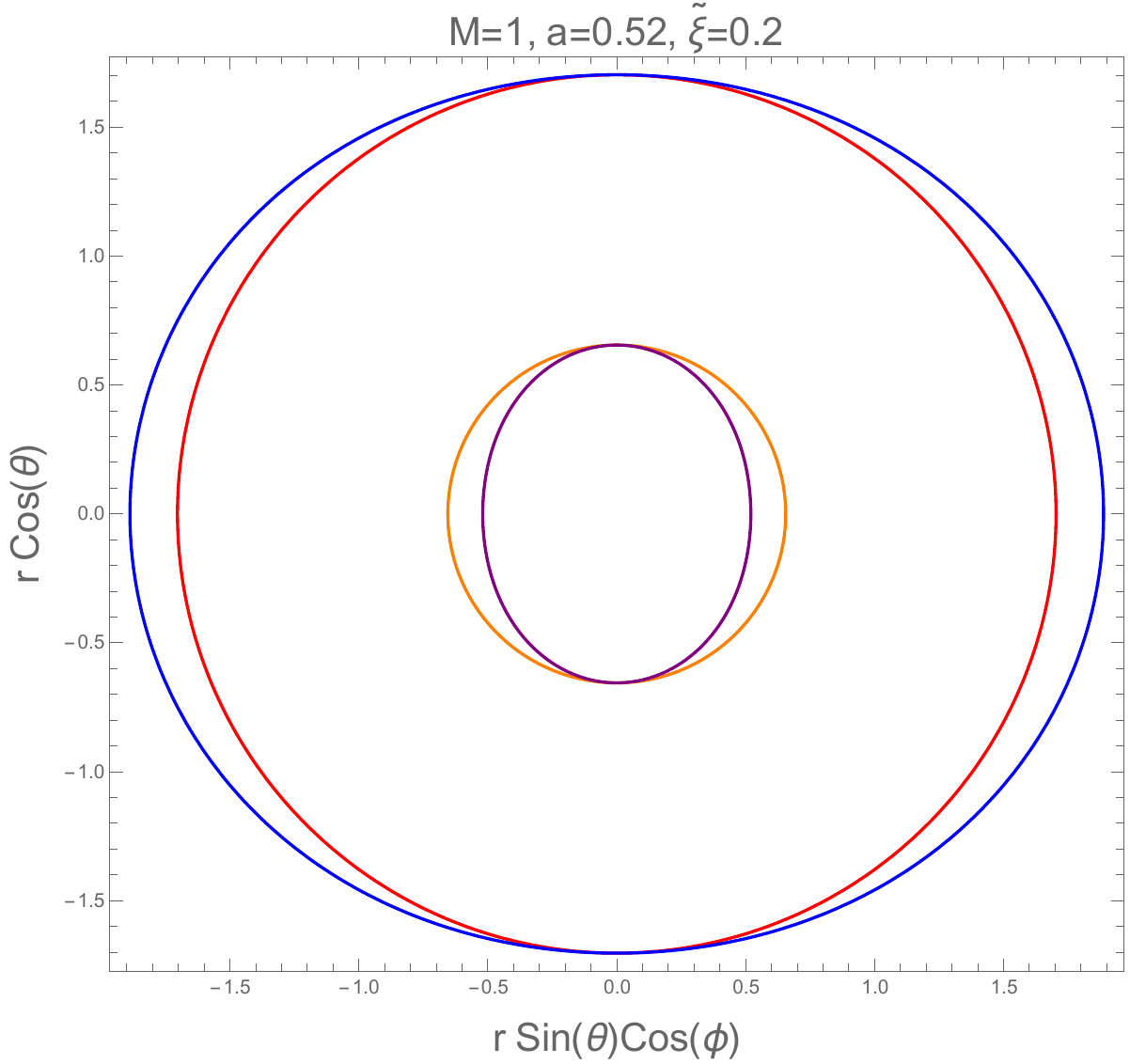}
       \includegraphics[width=0.22\linewidth]{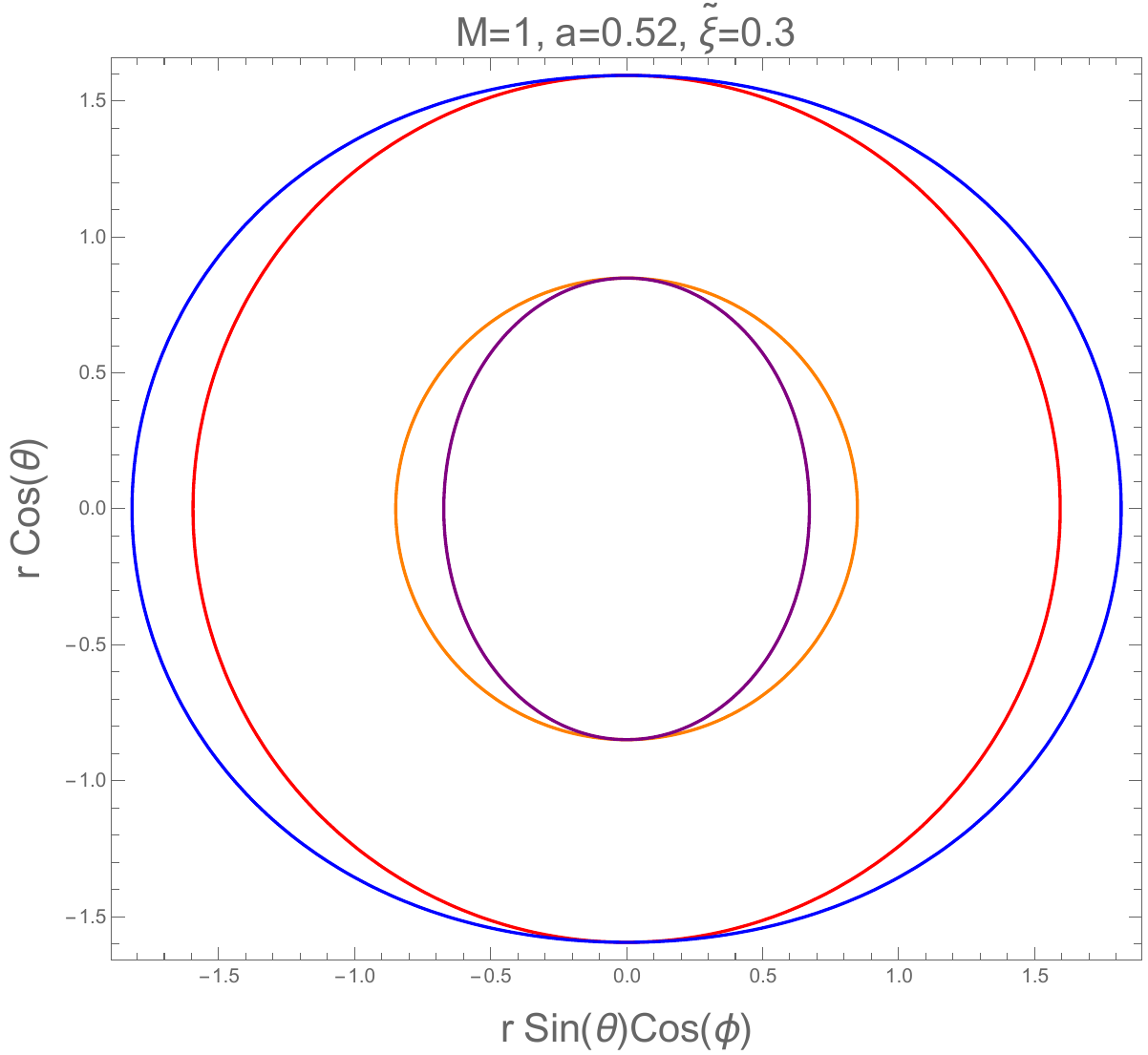}
        \includegraphics[width=0.22\linewidth]{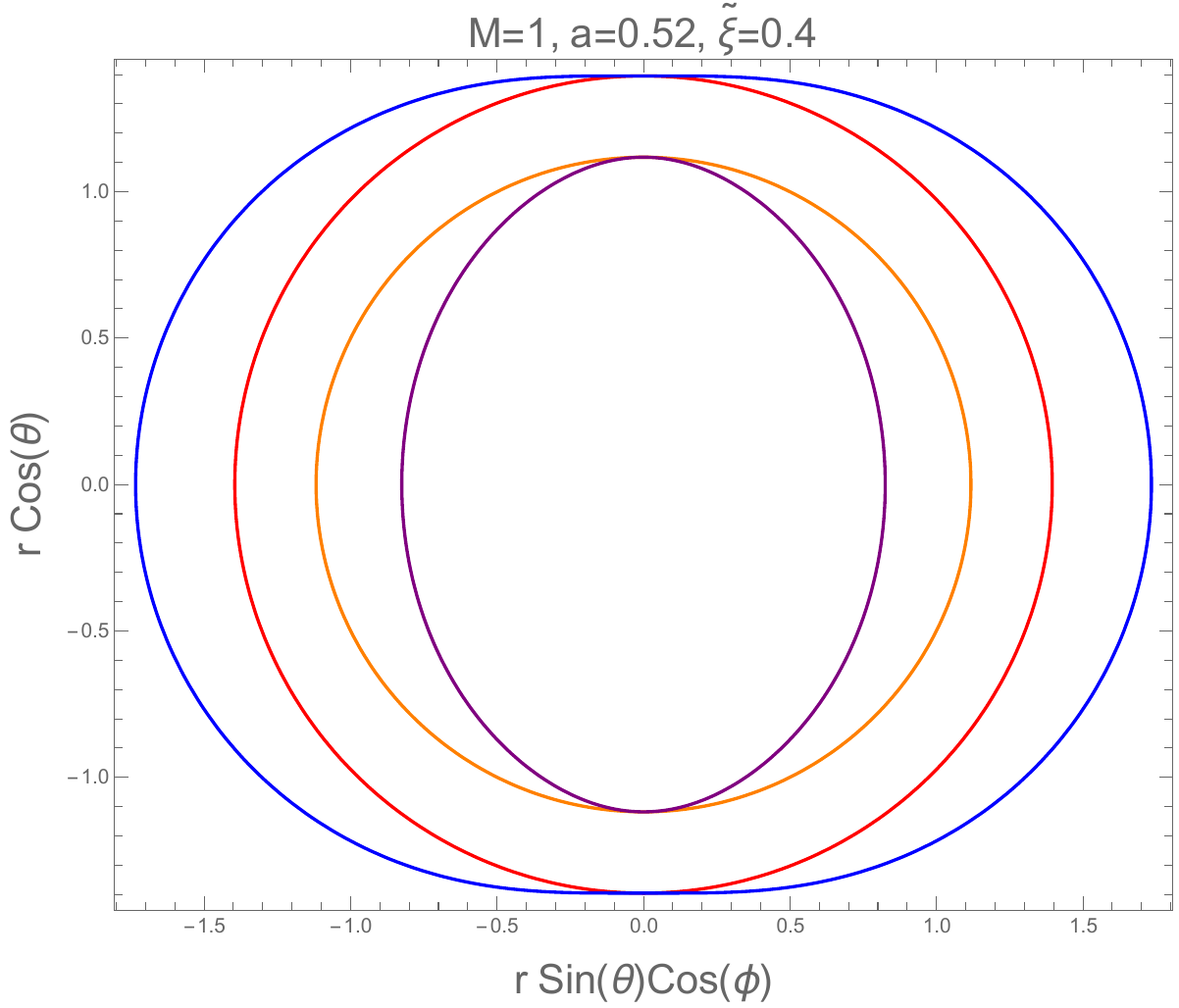}
         \includegraphics[width=0.22\linewidth]{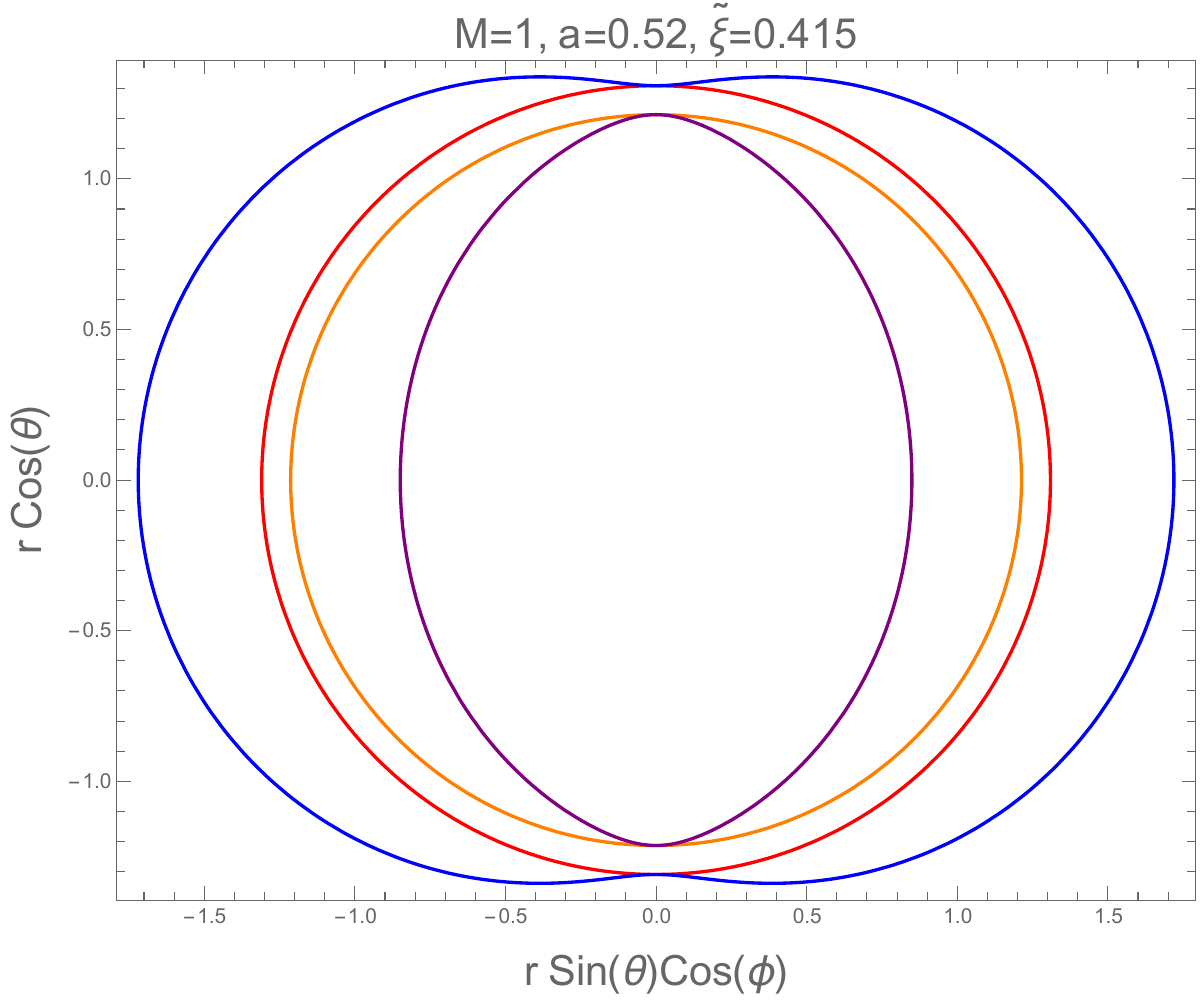}
          \includegraphics[width=0.22\linewidth]{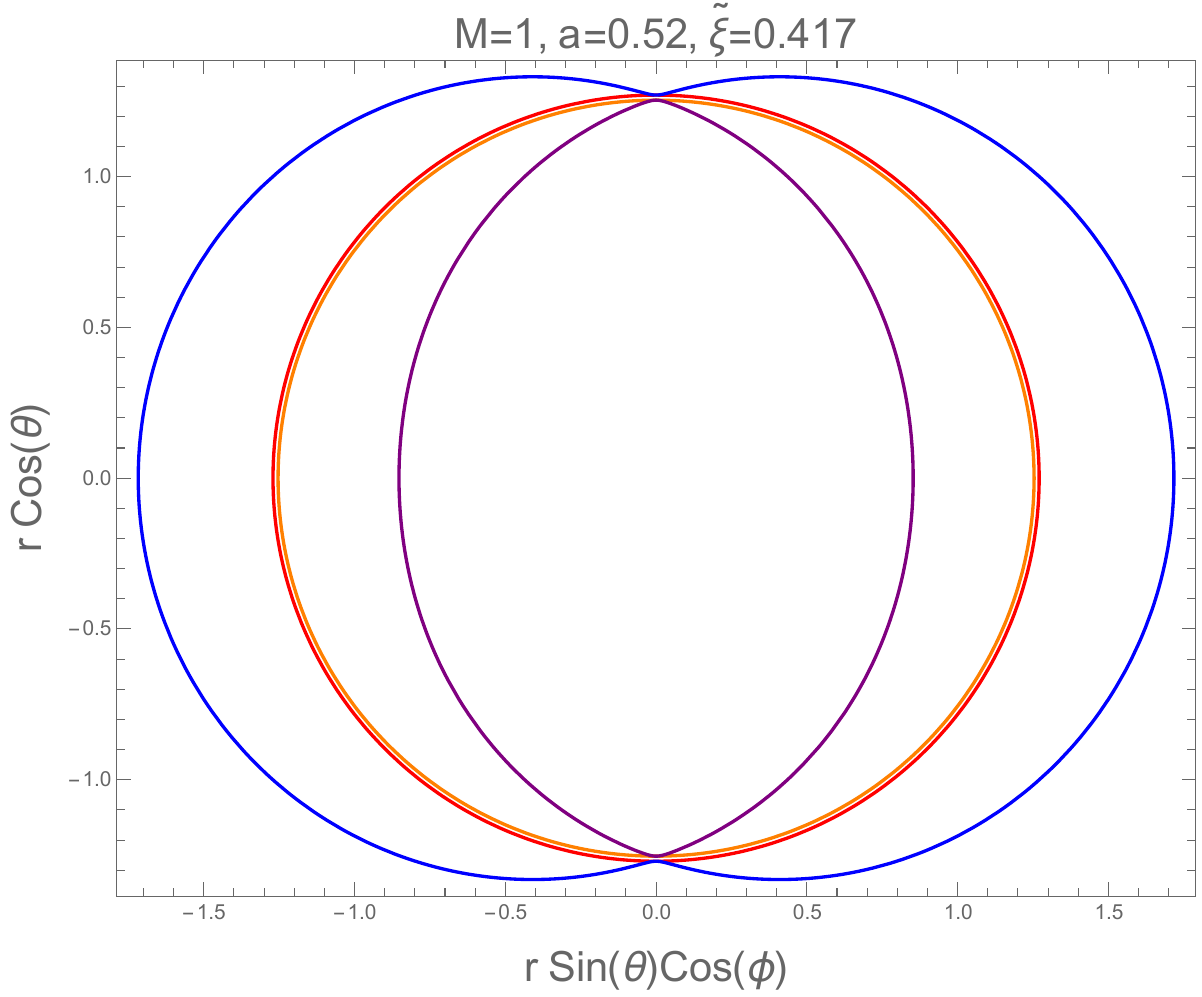}
           \includegraphics[width=0.22\linewidth]{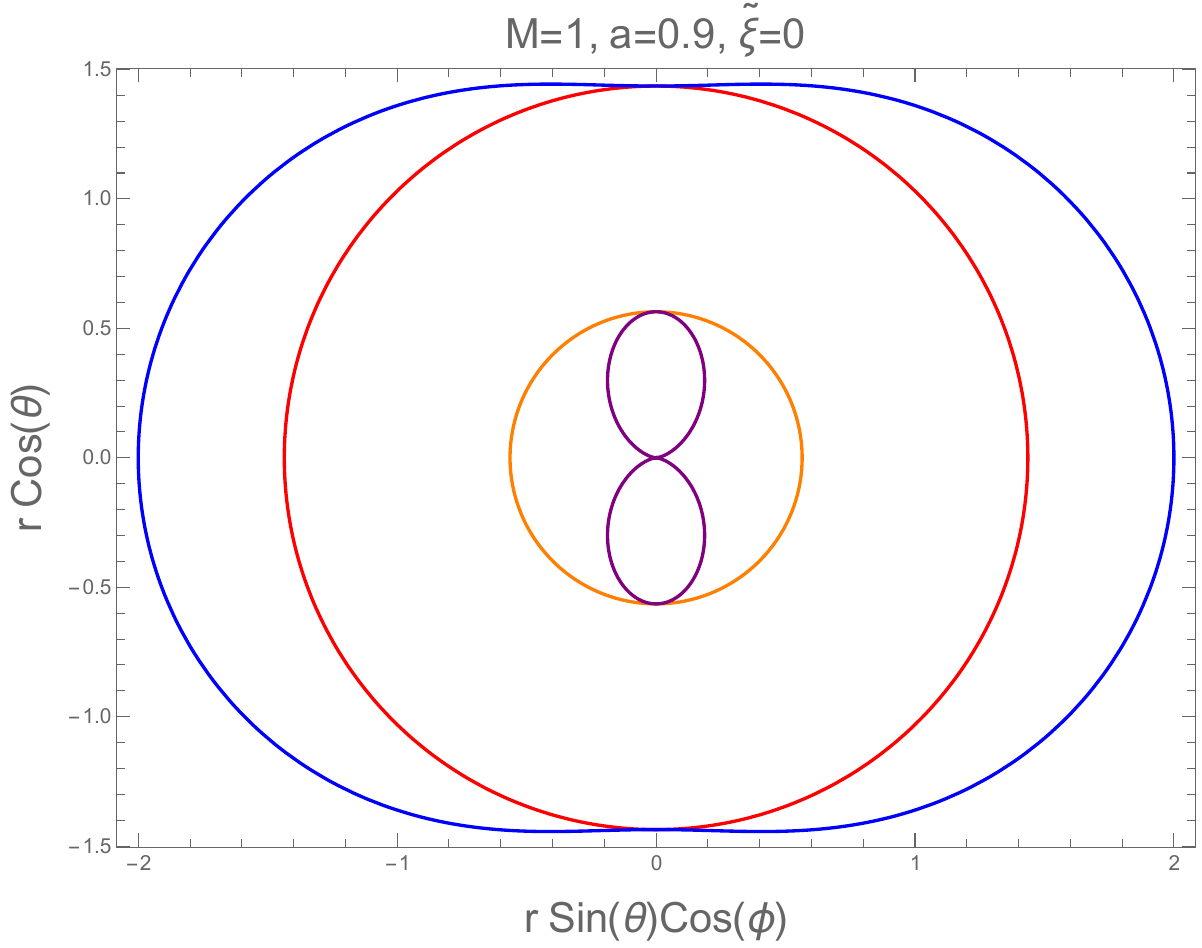}
            \includegraphics[width=0.22\linewidth]{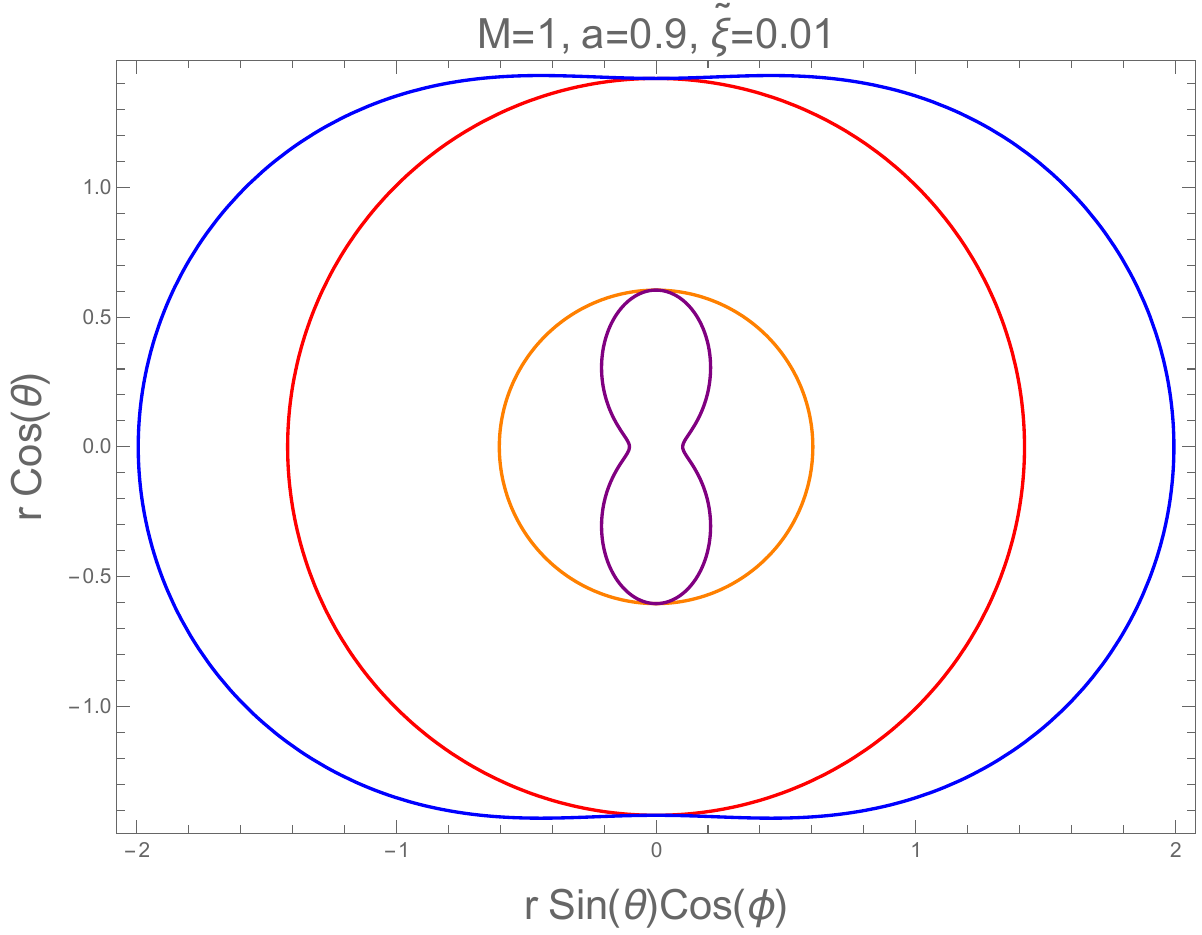}
             \includegraphics[width=0.22\linewidth]{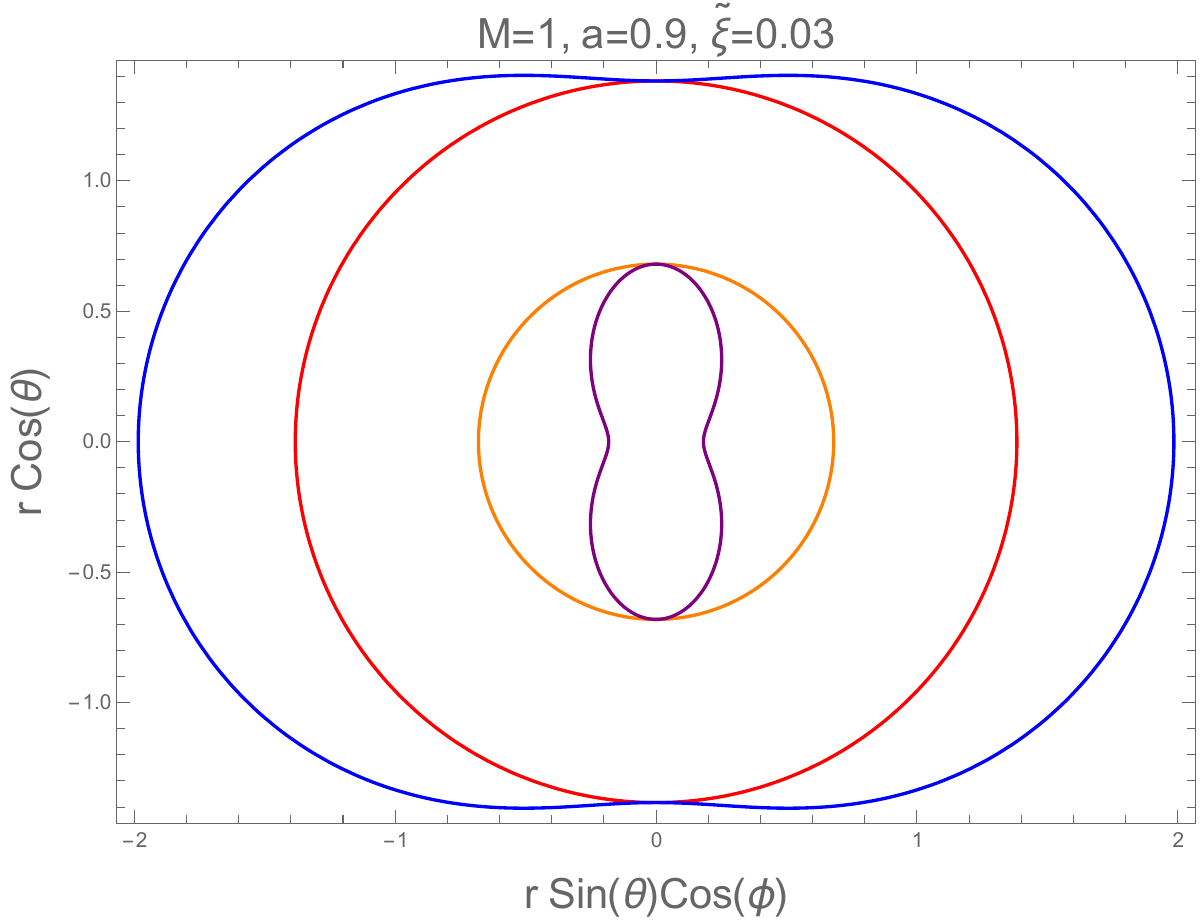}\quad~
              \includegraphics[width=0.22\linewidth]{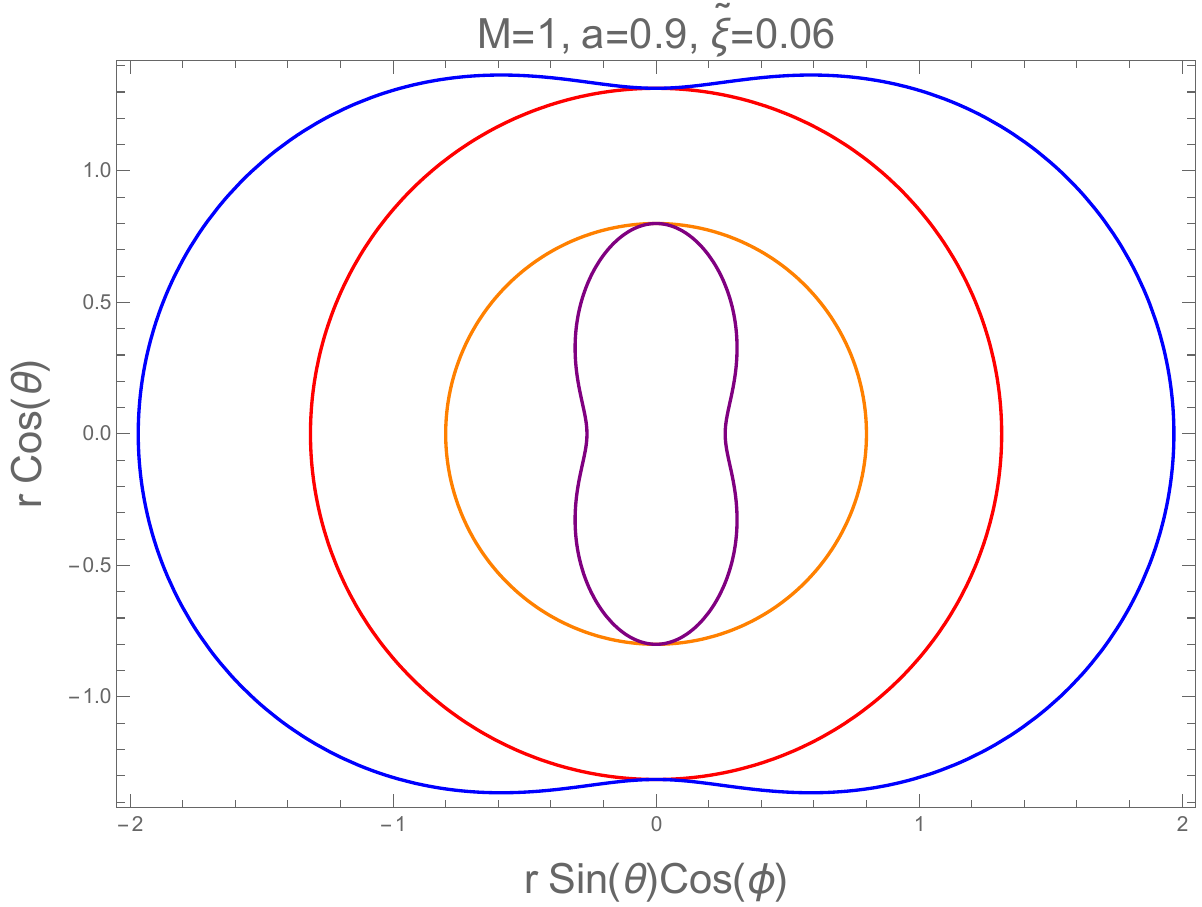}\quad~~
                 \includegraphics[width=0.22\linewidth]{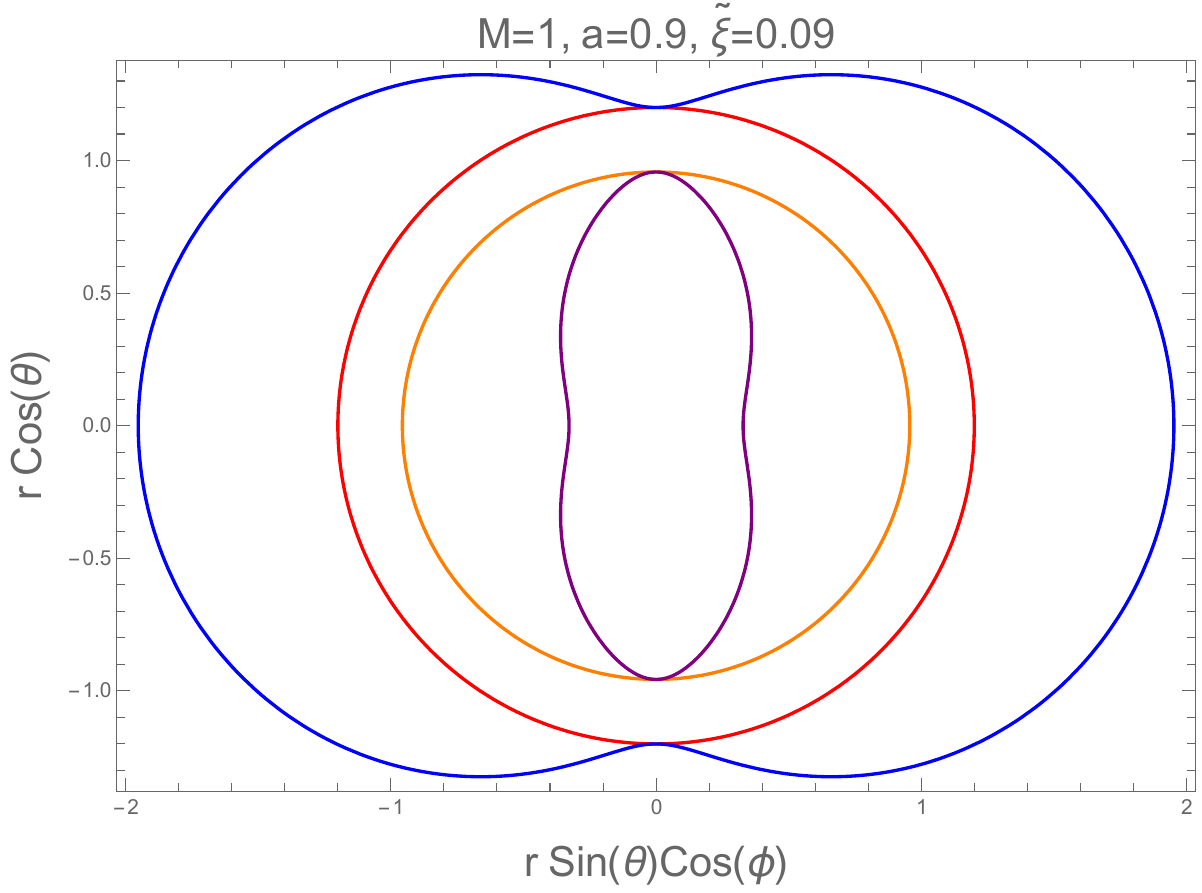}\quad~~
                   \includegraphics[width=0.22\linewidth]{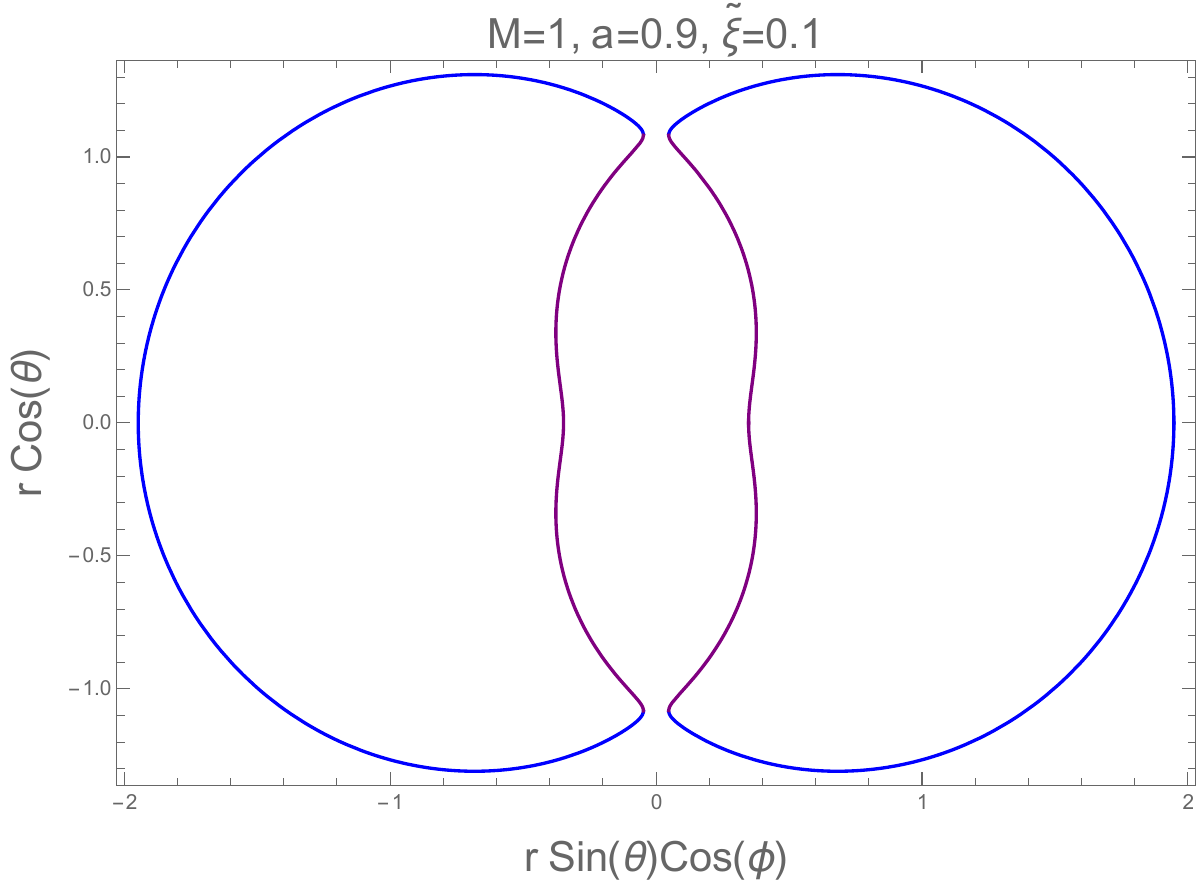}

 \caption{\scriptsize{Graphs showing change in shape of inner/outer horizons (Red/Orange) and inner/outer ergo-spheres (Blue/Purple) while the value of rotational parameter is $a=0.52, 0.9$. Note that ergo-sphere increases as $\tilde{\xi}$ increases.}} \label{FigEH}
\end{figure}

Modifications of the Kerr metric, that came as a result of the generalization of Lagrangian in the framework of running gravitational coupling,
also manifest in the properties of the black hole horizons. In the Boyer-Lindquist coordinates the event horizon, $r_{H}$, is given by the condition that at $r=r_{H}$ hypersurface is everywhere null, or $g^{rr}=0$. For the standard GR case it follows that the Kerr black hole will have two
solutions as long as $M>a$, while the case $M<a$ leads to existence of a naked singularity. This is however changed when one considers
the modifications coming from the spatial dependence of gravitational coupling. Now, the position of horizons is determined by the cubic equation
\begin{equation}
r^3 - 2Mr^{2}+a^2r+2M\tilde{\xi} =0.
\label{horizon}
\end{equation}
Although the horizon equation now has three solutions, only two of them can be positive - so there are no new horizons in this case. When compared to the horizons in the standard GR it can be checked that Eq. (\ref{horizon}) will tend to lead to
smaller separation between the inner and outer horizon. Moreover, the structure of black hole, described by its horizons, will now be changed and
will depend on the value of $\tilde{\xi}$ - determining if the horizons exist.\\ \\
For a general polynomial of order three, the number and type of roots is determined by its discriminant, $D$, so that for $D>0$ there exist three real solutions, in the
$D=0$ case the solutions are real and two of them are identical, while for $D<0$ one solution is real and two remaining ones are complex
conjugated. The existence of horizons, which should of course be real and positive, is then determined by $\tilde{\xi}_{c}$ for which $D=0$.
It follows that $\tilde{\xi}_{c}$ is given by
\begin{eqnarray}
\tilde{\xi}_c=\frac{-\left(9M^2a^2-8M^4\right)\pm\sqrt{M^2\left(4M^2-3a^2\right)^3}}{27M^2}.
\label{critical}
\end{eqnarray}
We note that $\tilde{\xi}_{c}$ will be physically viable (real valued) as long as the standard GR condition $M>a$ is satisfied.
Thus, for $\tilde{\xi}>\tilde{\xi}_c$ there will be no horizons, and this case leads to a naked singularity. For $\tilde{\xi}=\tilde{\xi}_c$ there will
be only one horizon (two identical positive roots), but this case is unstable since addition of some small amount of matter will violate
this condition. Finally, for $\tilde{\xi}<\tilde{\xi}_{c}$ the Kerr black hole will have one inner and one outer horizon. Since $\tilde{\xi}_{c}$
is given in general by theoretical consideration of asymptotic correction to the GR, this discussion constraints possible space of parameters for Kerr black hole that leads
to physically realistic solutions. \\ \\
The Killing horizon, being defined as the set of points where norm of the Killing vector becomes null, $K^{\mu}K_{\mu}=0$, in the case of running gravitational coupling is given by the solution of the following cubic equation
\begin{equation}
 r^3 - 2Mr^{2}+a^2 \cos^{2}(\theta) {r}+2M\tilde{\xi} =0.
\end{equation}
The discussion for event horizons structure depending on parameter $\tilde{\xi}$ given above can also be applied to the study of Killing horizons, with the replacement
\begin{align}
&\tilde{\xi}_{c} = \frac{-\left(9M^2a^2\cos ^{2}\theta -8M^4\right)\pm\sqrt{M^2\left(4M^2-3a^2\cos ^{2}\theta\right)^3}}{27M^2}.
\end{align}
The ergosphere, being the region between the outer event horizon and Killing horizon, is a region of particular interest since it is related to potentially observable processes related to the Kerr black hole, such as the extraction of energy via the Penrose process. It is therefore of special interest to investigate what are the effects of the IR gravity modifications on the ergosphere surface. It follows that the IR asymptotic safe modification typically increases the ergosphere region when compared to the one in the standard GR, as we show in Figure (\ref{Fig1}--\ref{FigEH}). It can be seen that the ergosphere surface tends to increase with the increase of parameter $\tilde{\xi}$. This in principle means that the region from where it is possible to extract energy from black hole by axial accretion of particles, via Penrose process, is bigger then in the GR for the equal parameters characterizing the black hole. However, the practical significance of this result is limited by the fact that $\tilde{\xi}$ needs to be a small parameter.

\subsection{Curvature Singularity}
An interesting characteristic of a black hole is its singularity, which can be defined mathematically when Kretschmann scalar $K$ tends to infinity. For metric given by Eq. (\ref{metric}) the Kretschmann scalar is
\begin{eqnarray}
K=\frac{M^2\mathcal{Z}(r,\theta,a,\tilde{\xi})}{8(r\Sigma)^6},
\end{eqnarray}
where
\begin{eqnarray*}\nonumber
\mathcal{Z}&=&384r^{12}-2560r^{10}\tilde{\xi}-a^8\tilde{\xi}^2+5888r^8\tilde{\xi}^2+4a^2\left(-1440r^{10}+4864r^8\tilde{\xi}+a^6\tilde{\xi}^2-1632r^6\tilde{\xi}^2\right)\cos^2\theta\\
&-&2a^4\left(-2880r^8+1280r^6\tilde{\xi}+3a^4\tilde{\xi}^2-64r^4\tilde{\xi}^2\right)\cos^4\theta+4a^6\left(-96r^6+a^2\tilde{\xi}^2+96r^2\tilde{\xi}^2\right)\cos^6\theta\\
&+&127a^8\tilde{\xi}^2\cos^8\theta+a^8\tilde{\xi}^2\sin^8\theta.
\end{eqnarray*}
We observe poles at $r=0$ and $\Sigma=r^2+a^2\cos^2\theta=0$ from where we interpret that the singularity exists at these points. The singularity associated with $\Sigma$  
constitutes a ring singularity analogous to that of Kerr black hole \cite{C}. However, we see that a new singularity, at $r=0$, has also 
appeared. This singularity can be traced to the form of IR correction to gravitational coupling (\ref{gtt}). It therefore follows that this 
singularity has no real physical significance since the approximation (\ref{gtt}) break in the region around $r=0$ and equations 
(\ref{gtt}) can no longer be applied. For a discussion of singularities around $r=0$ one would need to consider the UV limit of quantum corrections 
to gravitational coupling, which is beyond the scope of this work. 
\section{Geodesics Equations in Equatorial Plane}
This section is on equatorial geodesics of  rotating black hole solution including the effects of corrected 
gravitational coupling. The Lagrangian, for this metric, in the equatorial plane $(\theta=\frac{\pi}{2},\dot{\theta}=0)$ is written as \cite{C}
\begin{eqnarray}\nonumber
2\mathcal{L}=&-&\left[1-\frac{2M}{r}\left(1-\frac{\tilde{\xi}}{r^2}\right)\right]\dot{t}^2-\frac{4aM}{r}\left(1-\frac{\tilde{\xi}}{r^2}\right)\dot{t}\dot{\phi}\\\label{L1}
              &+&\frac{r^2}{\Delta}\dot{r}^2+\left[r^2+a^2+\frac{2a^2M}{r}\left(1-\frac{\tilde{\xi}}{r^2}\right)\right]\dot{\phi}^2.
\end{eqnarray}
The generalized momenta are given by
\begin{eqnarray}
p_t&=&-\left[1-\frac{2M}{r}\left(1-\frac{\tilde{\xi}}{r^2}\right)\right]\dot{t}-\frac{2aM}{r}\left(1-\frac{\tilde{\xi}}{r^2}\right)\dot{\phi}=-E,\label{pt}\\
p_{\phi}&=&-\frac{2aM}{r}\left(1-\frac{\tilde{\xi}}{r^2}\right)\dot{t}+\left[r^2+a^2+\frac{2a^2M}{r}\left(1-\frac{\tilde{\xi}}{r^2}\right)\right]\dot{\phi}=L,\label{pf}\\
p_{r}&=&\frac{r^2}{\Delta}\dot{r}\label{pr},
\end{eqnarray}
where dots over $r$, $t$ and $\phi$ denote derivatives with respect to affine parameter $\tau$. It can be easily seen that Lagrangian does not depend on $t$ and $\phi$, therefore $p_t$ and $p_\phi$ are conserved quantities.\\
The Hamiltonian is given by
\begin{eqnarray}
\mathcal{H}=p_t\dot{t}+p_\phi\dot{\phi}+p_r\dot{r}-\mathcal{L}.
\end{eqnarray}
It takes the form
\begin{eqnarray}
2\mathcal{H}&=&\left[-\left(1-\frac{2M}{r}\left(1-\frac{\tilde{\xi}}{r^2}\right)\right)\dot{t}-\frac{2aM}{r}\left(1-\frac{\dot{\xi}}{r^2}\right)\dot{\phi}\right]\dot{t}\\\nonumber &+&\left[-\frac{2aM}{r}\left(1-\frac{\tilde{\xi}}{r^2}\right)\dot{t}+\left(r^2+a^2+\frac{2a^2M}{r}\left(1-\frac{\tilde{\xi}}{r^2}\right)\dot{\phi}\right)\right]\dot{\phi}+\frac{r^2}{\Delta}\dot{r}^2,\\
2\mathcal{H}&=&{-E}\dot{t}+L\dot{\phi}+\frac{r^2}{\Delta}\dot{r}=\delta=constant,\label{H1}
\end{eqnarray}
where Hamiltonian is constant as it is $t$ independent and $\delta=-1,0,1$ gives timelike, null and spacelike geodesics respectively. Solving Eq. (\ref{pt}) and Eq. (\ref{pf}) yield:
\begin{eqnarray}
\dot{t}&=&\frac{1}{\Delta}\left[\left(r^2+a^2+\frac{2a^2M}{r}\left(1-\frac{\tilde{\xi}}{r^2}\right)\right)E-\frac{2aM}{r}\left(1-\frac{\tilde{\xi}}{r^2}\right)L\right],\label{t1}\\
\dot{\phi}&=&\frac{1}{\Delta}\left[\frac{2aM}{r}\left(1-\frac{\tilde{\xi}}{r^2}\right)E+\left(1-\frac{2M}{r}\left(1-\frac{\tilde{\xi}}{r^2}\right)\right)L\right].\label{f1}
\end{eqnarray}
On substituting Eq. (\ref{t1}) and Eq. (\ref{f1}) in Eq. (\ref{H1}), we get the radial equation of motion
\begin{eqnarray}
r^2\dot{r}^2=\Delta\delta+r^2E^2+\frac{2M}{r}\left(1-\frac{\tilde{\xi}}{r^2}\right)\left(aE-L\right)^2+\left(a^2E^2-L^2\right).\label{r11}
\end{eqnarray}
In the limit $\tilde{\xi}\rightarrow{0}$, Eq. (\ref{r11}) takes the form of radial equation in the Kerr black hole case.
\subsection{Null geodesics}
In equatorial plane, the null geodesics are rendered when $\delta$ gets zero in Eq. (\ref{r11}), which then becomes
\begin{eqnarray}
r^2\dot{r}^2=r^2E^2+\frac{2M}{r}\left(1-\frac{\tilde{\xi}}{r^2}\right)\left(aE-L\right)^2+\left(a^2E^2-L^2\right).\label{r12}
\end{eqnarray}
For convenience, introduce an impact parameter $D=L/E$ in Eq. (\ref{r12}). Through this parameter the angular momentum can be expressed in terms of energy. Two cases may arise here: either $D=a$ or $D\neq{a}$.
\subsubsection{CASE (I); when D=a}
As a particular case, consider $D=a$ or $L=aE$. As a result of which Eq. (\ref{t1}), Eq. (\ref{f1}) and Eq. (\ref{r12}) imply
\begin{eqnarray}
\dot{t}&=&\frac{r^2+a^2}{\Delta}E,\label{t2}\\
\dot{\phi}&=&\frac{aE}{\Delta},\label{f2}\\
\dot{r}&=&\pm{E}.\label{r2}
\end{eqnarray}
Notice here that when $\Delta=0$ (at horizon), both $\dot{t}$ and $\dot{\phi}$ go to infinity. This implies that $t$ and $\phi$ operate as `bad coordinates' in the vicinity of horizon, but this singularity vanishes in Eq. (\ref{r2}), the expression for $\dot{r}$.
Using above equations, the differentials of $t$ and $\phi$, with respect to $r$, are computed as
\begin{eqnarray}
\frac{dt}{dr}&=&\pm\frac{\left(r^2+a^2\right)}{\Delta},\label{r21}\\
\frac{d\phi}{dr}&=&\pm\frac{a}{\Delta},\label{r22}
\end{eqnarray}
where $+$ and $-$ signs in Eq.s (\ref{r2}-\ref{r22}) stand respectively for the trajectory of outgoing and ingoing photon.
\begin{figure}[H]
    \includegraphics[width=0.45\linewidth]{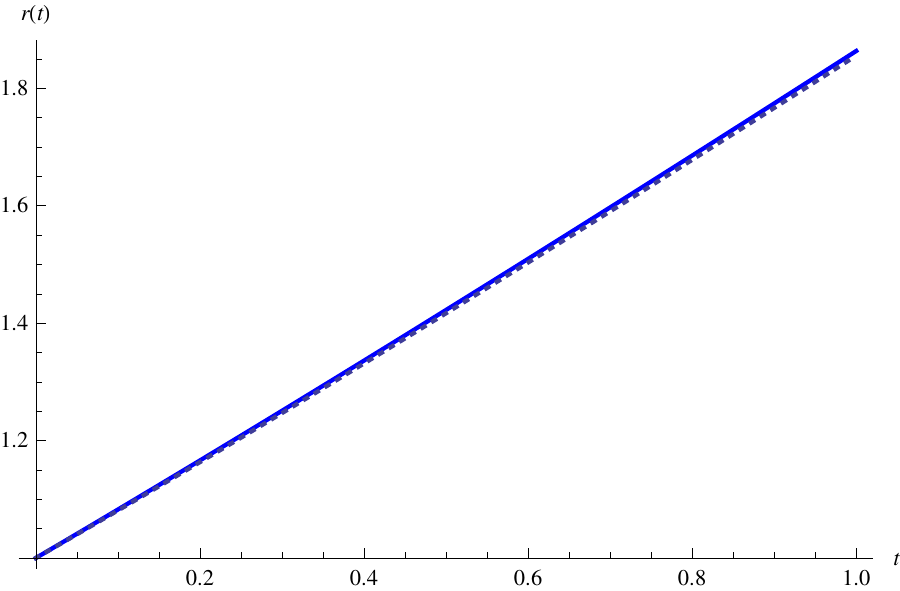}
    \includegraphics[width=0.45\linewidth]{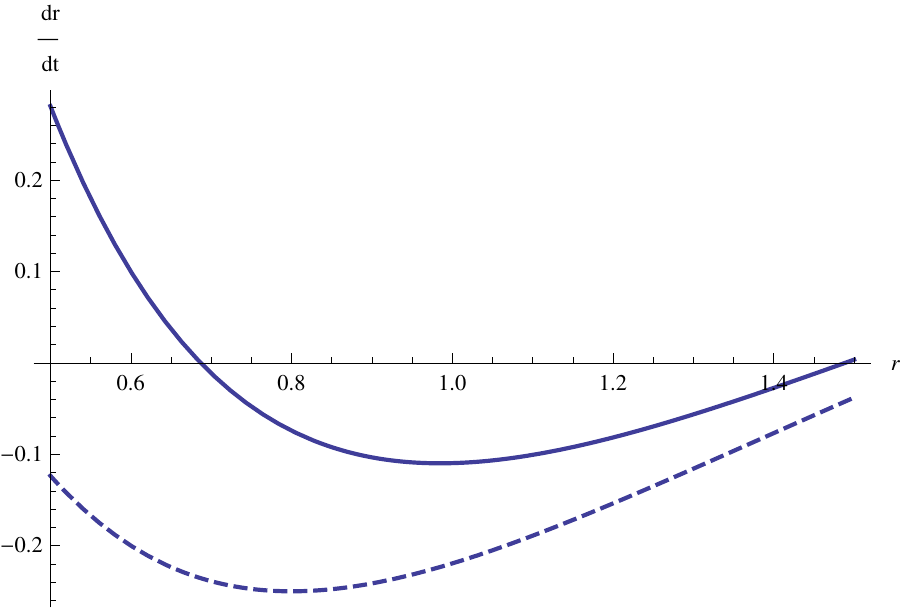}
    \caption{\scriptsize{The left panel represents the outgoing trajectory of photons with respect to time $t$. The solid line represents when $\tilde{\xi}=0.09$ and dashed line is when $\tilde{\xi}=0$ i.e Kerr case. The values of parameters are $a=0.1$ and $M=0.1$ with the initial condition $r(0)=10 M$. On the right panel the phase portret is depicted for the same parameters, where again the solid line represents the case when $\tilde{\xi}=0.09$ and dashed line is when $\tilde{\xi}=0$. The fixed points are getting closer in the asymptotically safe gravity, as opposed to the GR case where the fixed points are at maximum distance.}} \label{Figrt}
\end{figure}
\begin{figure}[H]
    \includegraphics[width=0.45\linewidth]{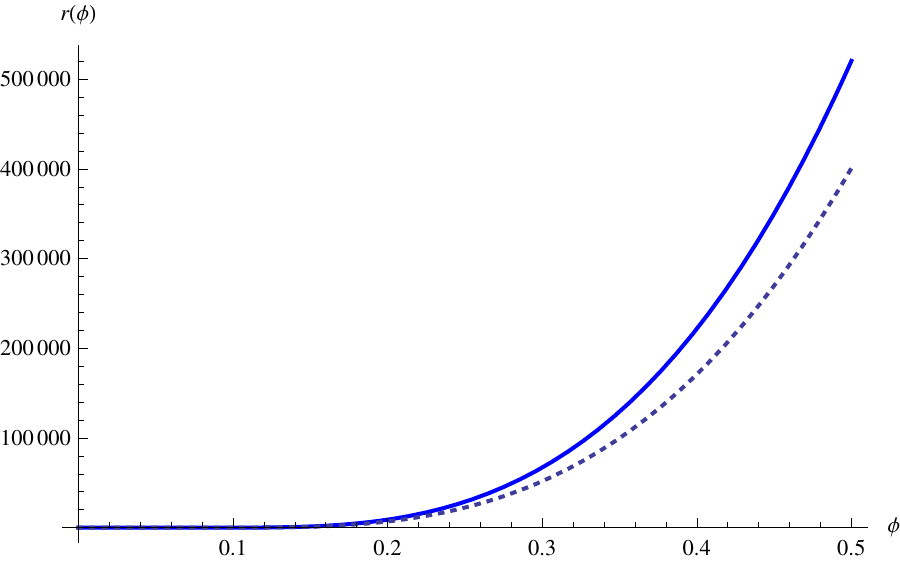}
    \includegraphics[width=0.45\linewidth]{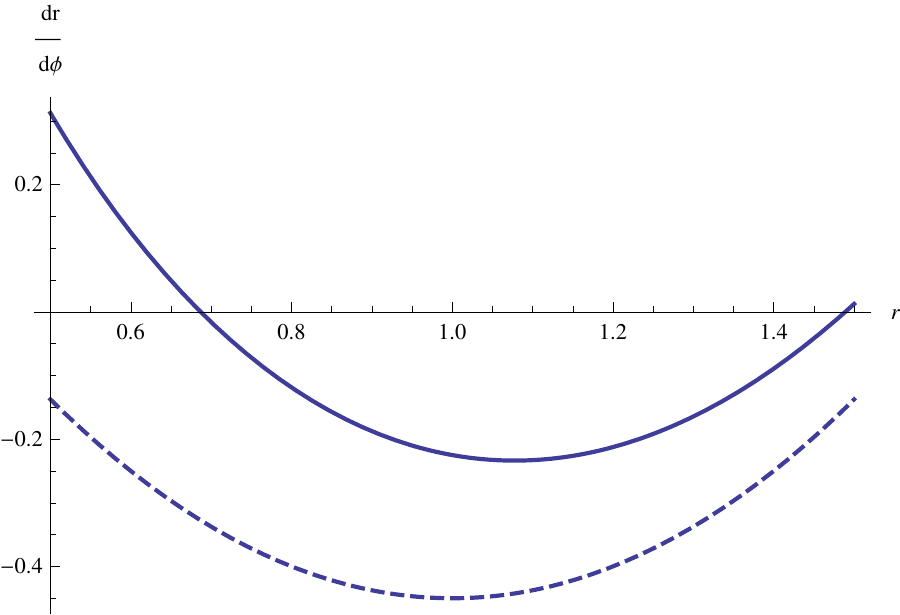}
    \caption{\scriptsize{The left panel represents the outgoing trajectory of photons with respect to angle $\phi$. The solid line represents when $\tilde{\xi}=0.09$ and dashed line is when $\tilde{\xi}=0$ i.e Kerr case for the values of parameters $a=0.1$ and $M=0.1$ with the initial condition $r(0)=10 M$. On the right panel the phase portrait is depicted for the same parameters, where again the solid line represents the case when $\tilde{\xi}=0.09$ and dashed line is the GR case $\tilde{\xi}=0$. The fixed points are again getting closer in the asymptotically safe gravity, in the contrast to the GR case where the distance between them is maximal.
   }}\label{Figrphi}
\end{figure}
\begin{figure}[H]
    \includegraphics[width=0.45\linewidth]{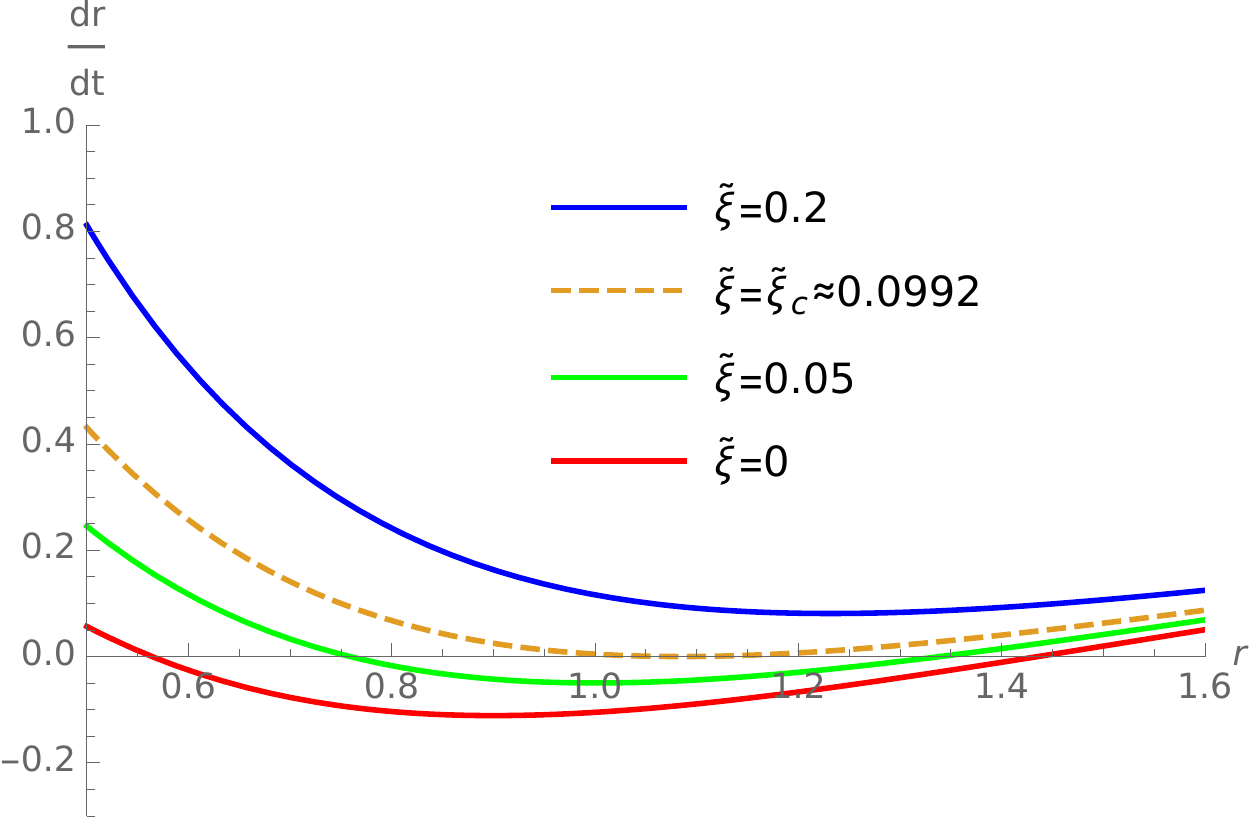}
     \includegraphics[width=0.45\linewidth]{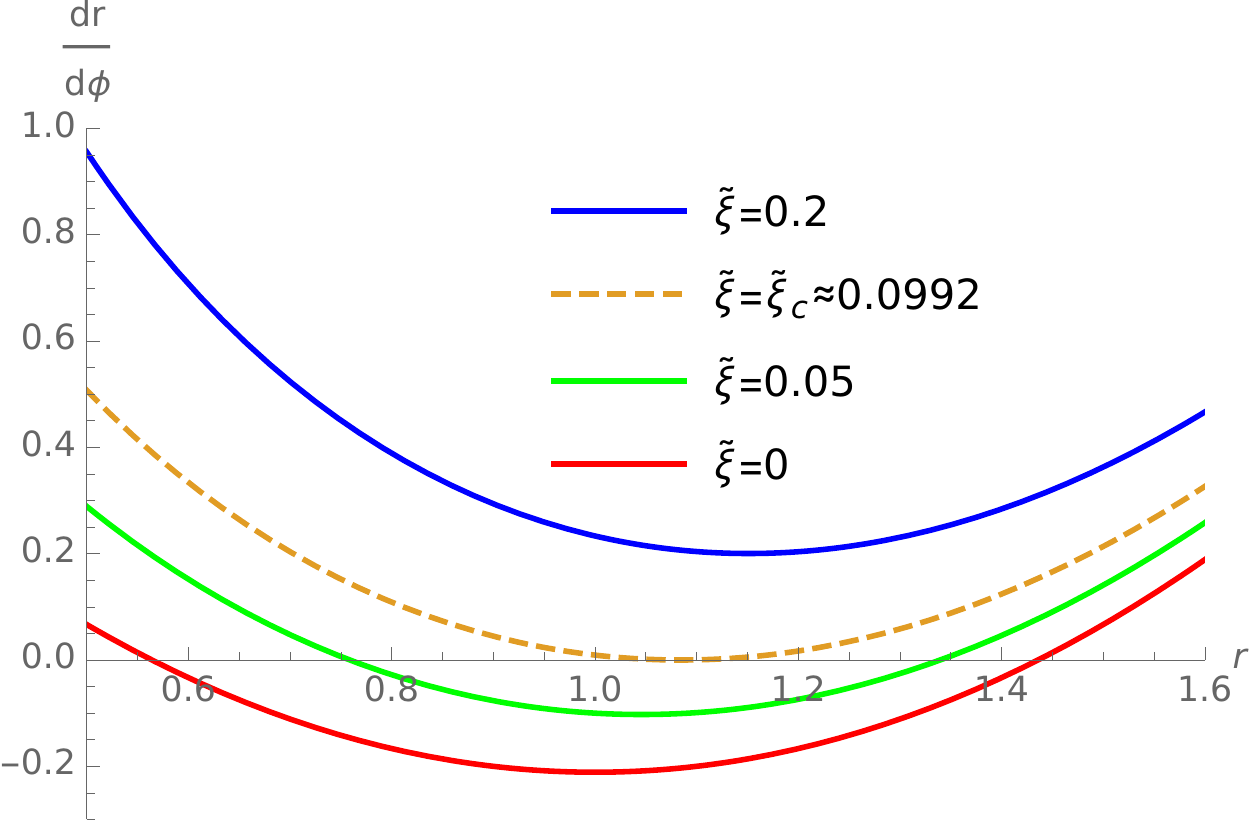}
        \caption{\scriptsize{Phase space portrait, $dr/dt$ on the left panel and $dr/d\phi$ on the right, for different values of $\tilde{\xi}$, but for fixed parameters $M=1$ and $a=0.9$. Clearly, the critical value for $\tilde{\xi}=\tilde{\xi}_{c}$ represent the bifurcation point for which the fixed points do not exist anymore.}}
        \label{Figbif}
\end{figure}
The trajectory for outgoing photon is numerically solved and plotted in Figure (\ref{Figrt}) and Figure (\ref{Figrphi}) where the initial condition $r(0)=10 M$ is imposed. It can be seen that there are no high deviation from the GR counterpart in the trajectory $r(t)$ as expected from small $\tilde{\xi}$, but from the phase portrait, when $d{r}/dt$ is considered as a function of $r$, the fixed points are getting closer to each other by increasing $\tilde{\xi}$. Same are the results for the photon trajectory with respect to angle $\phi$. To get a qualitative description of equations (\ref{r21}) and (\ref{r22}) one
can also easily analyze them in the phase space for different values of $\tilde{\xi}$. It can be seen that there
exists a bifurcation point for which the two real fixed points
vanish, and for this parameters the solution leads to a naked singularity, the phase spaces are plotted in Figure (\ref{Figbif}).
\subsubsection{CASE(II); when $D\neq{a}$}
As a general case consider $D\neq{a}$, which consequently gives circular orbit $r=r_c$ of photon. Introduce an impact parameter $D_c=L_c/E_c$. The radial equation (\ref{r12}) along with its derivative takes the following form
\begin{eqnarray}
r_c^2+\frac{2M}{r_c}\left(1-\frac{\tilde{\xi}}{r_c^2}\right)\left(a-D_c\right)^2+\left(a^2-D_c^2\right)=0,\\
r_c-\frac{M}{r_c^2}\left(1-\frac{3\tilde{\xi}}{r_c^2}\right)\left(a-D_c\right)^2=0.
\end{eqnarray}
Combining above two equations implies the following result
\begin{eqnarray}
r_c^2-3Mr_c+\frac{5M\tilde{\xi}}{r_c}\pm{2a\sqrt{Mr_c\left(1-\frac{3\tilde{\xi}}{r_c^2}\right)}}=0.
\end{eqnarray}
The real positive solution of the above equation will give circular photon orbit. For $\tilde{\xi}=0$, it matches with circular photon orbit for Kerr black hole.
\subsection{Time-like Geodesics}
To investigate time-like geodesics, take $\delta=-1$. Notice that equations for $\dot{\phi}$ and $\dot{t}$ remain unchanged, while Eq. (\ref{r11}) becomes
\begin{eqnarray}\label{g1}
r^2\dot{r}^2=-\Delta+r^2E^2+\frac{2M}{r}\left(1-\frac{\tilde{\xi}}{r^2}\right)\left(aE-L\right)^2+\left(a^2E^2-L^2\right),
\end{eqnarray}
where $E$ is now described as the energy per unit mass of the particle moving in a trajectory. Two cases arise here, either $L=aE$, a special case, or $L\ne{aE}$, a general case which can lead us to circular and associated orbits.
\subsubsection{Special Case: when $L=aE$}
\begin{figure}[H]
    \includegraphics[width=0.45\linewidth]{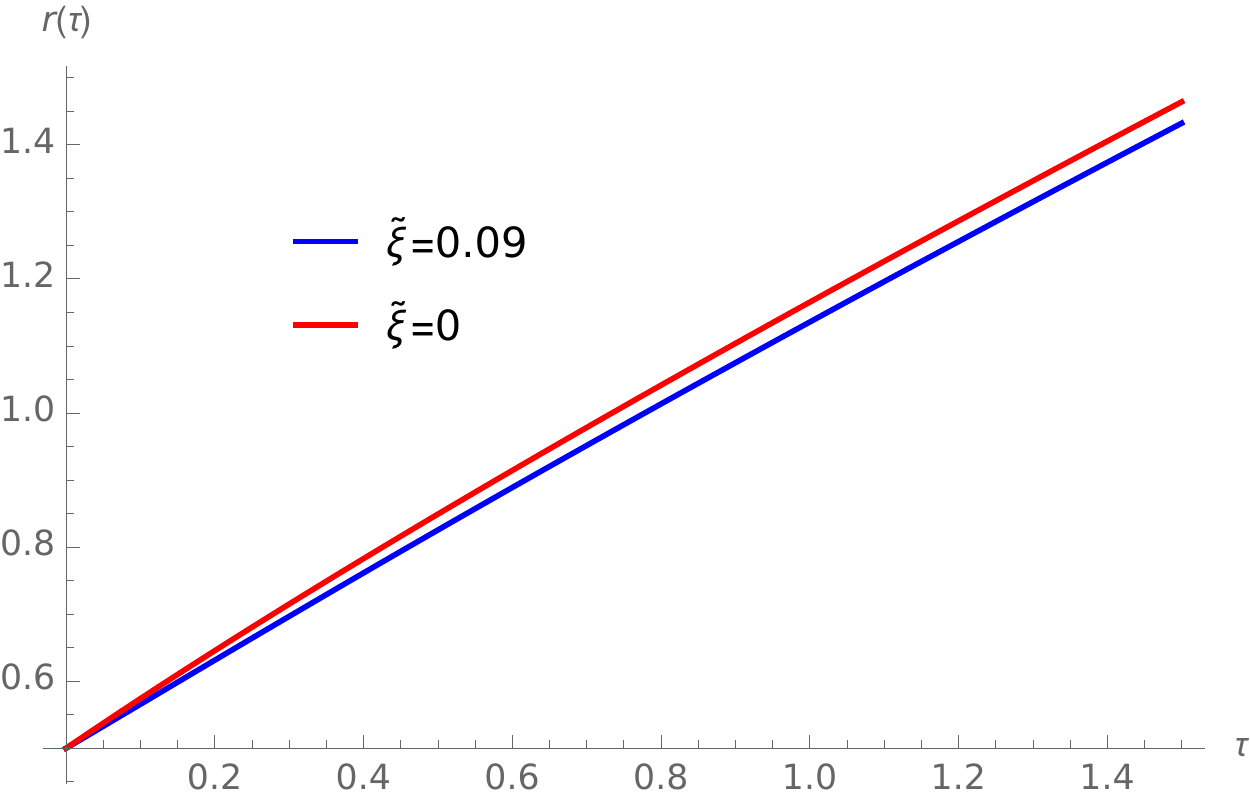}
     \includegraphics[width=0.45\linewidth]{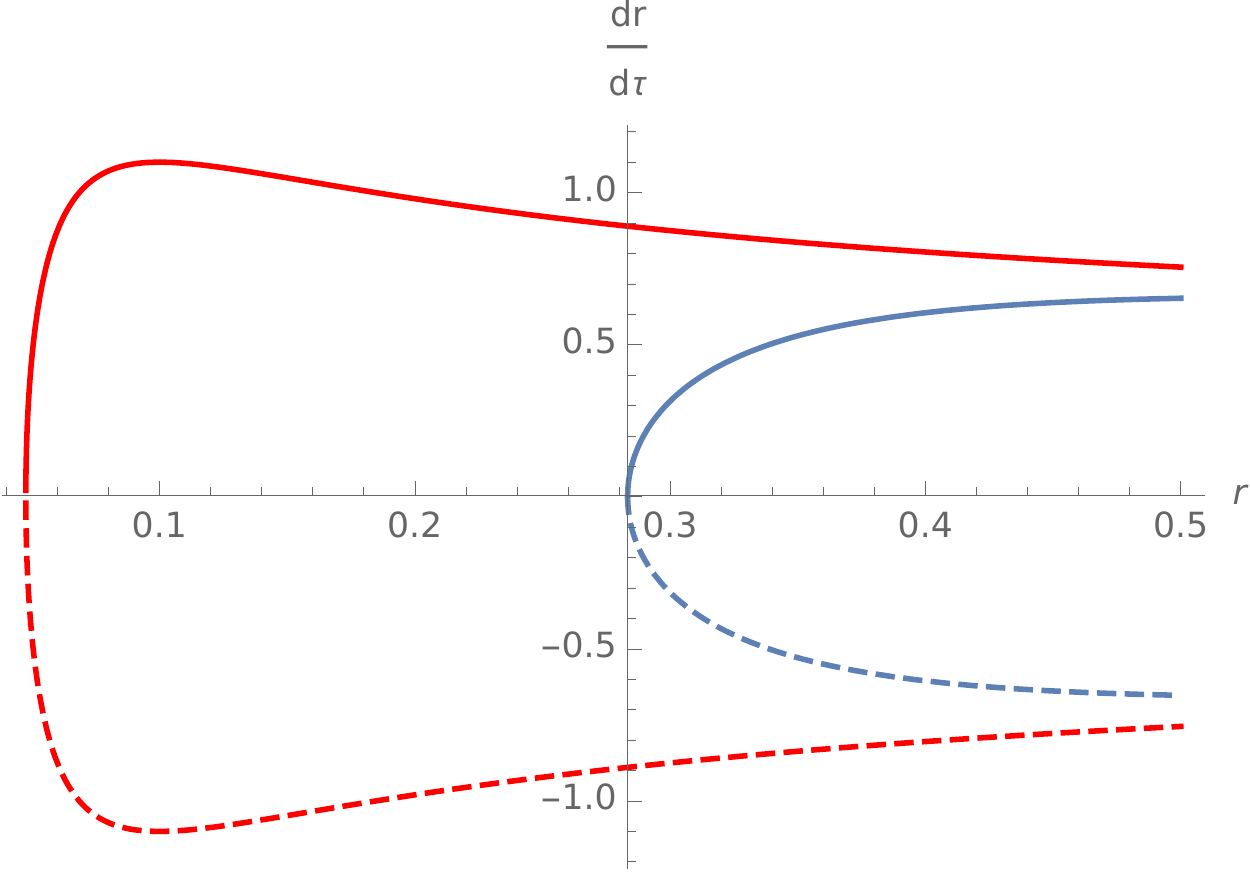}
    \caption{{The left panel represents the trajectory of a particle $r(t)$, where the blue line represents the asymptotically safe gravity case with $\tilde{\xi}=0.09$ and the red line represents the GR case $\tilde{\xi}=0$ i.e Kerr case. The values of parameters are $a=0.1$, $M=0.1$ and $E=1.1$ with the initial condition $r(0)=10M$. On the right panel the phase portrait is depicted for the same parameters, where the blue line is the phase diagram for $\tilde{\xi}=0.09$ and the red line is the GR
    case where $\tilde{\xi}=0$,
    the solid line represents the positive square root and dashed line the negative for each case. The fixed point is increased in the asymptotically safe gravity $(r_{*}\approx0.285)$ from the GR case ($r_{*}\approx0.05$).
   }}\label{Figrtau}
\end{figure}

Consider $L=aE$, Eq. (\ref{g1}) gives
\begin{eqnarray}\label{g2}
r^2\dot{r}^2=r^2\left(E^2-1\right)+2Mr\left(1-\frac{\tilde{\xi}}{r^2}\right)-a^2,
\end{eqnarray}
while $\dot{t}$ and $\dot{\phi}$ are the same as for null geodesics. Integrate Eq. (\ref{g2})
\begin{eqnarray}\label{tau11}
\tau=\int\frac{rdr}{\sqrt{r^2\left(E^2-1\right)+2Mr\left(1-\frac{\tilde{\xi}}{r^2}\right)-a^2}}.
\end{eqnarray}
The above equation is somewhat hideous to solve analytically. Its numerical solution is plotted in Figure (\ref{Figrtau}). Again, as there are no high deviations from the GR case it could be more interesting to analyze the phase portrait for each case. It can be seen that the fixed point is higher if the running gravitational coupling is considered, compared to the
GR counterpart. Also the phase space shows higher deviations near the fixed point but asymptotically as $r\rightarrow \infty$ the phase spaces coincide in the two cases. The phase space diagram is also plotted in Figure (\ref{Figrtau}) but on the right panel.
\subsubsection{General Case: when $L\neq{aE}$}
To investigate the general case, again take into account the radial Eq. (\ref{g1}). By introducing the reciprocal radius $u=1/r$, the equation takes the form
\begin{eqnarray}\nonumber
\mathcal{F}(u)=u^{-4}\dot{u}^2&=&E^2+2Mu^3\left(1-\tilde{\xi}u^2\right)\left(aE-L\right)^2+\left(a^2E^2-L^2\right)u^2\\\label{g3}
&-&\left(1-2Mu\left(1-\tilde{\xi}u^2\right)+a^2u^2\right),
\end{eqnarray}
where $u$ is the independent variable.\\
The task now is to compute the values of $E$ and $L$ for the circular orbit at the reciprocal radius $u={1}/{r}$. Circular orbits exist when $\mathcal{F}(u)=0$ and $\mathcal{F}^\prime(u)=0$. Also, assume $x=L-aE$ in Eq. (\ref{g3}), to get
\begin{eqnarray}\label{g4}
E^2-\left(1-2Mu\left(1-\tilde{\xi}u^2\right)+a^2u^2\right)+2Mu^3\left(1-\tilde{\xi}u^2\right)x^2-\left(x^2+2xaE\right)u^2=0,
\end{eqnarray}
\begin{eqnarray}\label{g5}
M\left(1-3\tilde{\xi}u^2\right)-a^2u+3Mu^2x^2\left(1-\frac{5}{3}\tilde{\xi}u^2\right)-\left(x^2+2xaE\right)u=0.
\end{eqnarray}
Solve Eq. (\ref{g4}) and Eq. (\ref{g5}), to reach to the following form
\begin{eqnarray}\label{g6}
E^2=Mu^3x^2\left(1-3\tilde{\xi}u^2\right)+1-Mu\left(1+\tilde{\xi}u^2\right),
\end{eqnarray}
\begin{eqnarray}\label{g7}
2xaEu=3Mu^2x^2\left(1-\frac{5}{3}\tilde{\xi}u^2\right)+M\left(1-3\tilde{\xi}u^2\right)-x^2u-a^2u.
\end{eqnarray}
Using Eq. (\ref{g6}) and Eq. (\ref{g7}), $E$ is eradicated and the quadratic equation in $x$ is obtained as
\begin{eqnarray}\nonumber
&x^4&u^2\left[3Mu\left(1-\frac{5}{3}\tilde{\xi}u^2-1\right)^2-4Ma^2u^3\left(1-3\tilde{\xi}u^2\right)\right]\\\nonumber
&-&2x^2u\left[\Big(3Mu(1-\frac{5}{3}\tilde{\xi}u^2)-1\Big)\Big(a^2u-M(1-3\tilde{\xi}u^2)+2a^2u(1-Mu(1+\tilde{\xi}u^2))\Big)\right]\\\label{g8}
&+&\Bigg[a^2-M(1-3\tilde{\xi}u^2)\Bigg]^2=0.
\end{eqnarray}
The discriminant of the above equation is given by
\begin{eqnarray}
\mathcal{D}=16a^2Mu^3{\Delta_u}^2,
\end{eqnarray}
where $\Delta_u=1+a^2u^2-2Mu\left(1-\tilde{\xi}u^2\right)$. The calculations can be eased by considering
\begin{eqnarray}\nonumber
1-3Mu\left(1-\frac{5}{3}\tilde{\xi}u^2\right)-4Ma^2u^3\left(1-3\tilde{\xi}u^2\right)=\mathcal{F}_+\mathcal{F}_-,
\end{eqnarray}
where
\begin{eqnarray}\nonumber
\mathcal{F}_\pm=1-3Mu\left(1-\frac{5}{3}\tilde{\xi}u^2\right)\pm{2a\sqrt{Mu^3\left(1-3\tilde{\xi}u^2\right)}}.
\end{eqnarray}
The solution of Eq. (\ref{g8}) is then simply computed as
\begin{eqnarray}\label{g9}
x^2u^2=\frac{\Delta_u\mathcal{F}_{\pm}-\mathcal{F}_+\mathcal{F}_-}{\mathcal{F}_+\mathcal{F}_-}=\frac{\Delta_u-\mathcal{F}_{\mp}}{\mathcal{F}_{\mp}},
\end{eqnarray}
where
\begin{eqnarray}\nonumber
\Delta_u-\mathcal{F}_\mp=u\left[a\sqrt{u}\pm\sqrt{M\left(1-3\tilde{\xi}u^2\right)}\right]^2.
\end{eqnarray}
Finally, we get
\begin{eqnarray}\label{g9}
x=-\frac{a\sqrt{u}\pm\sqrt{M\left(1-3\tilde{\xi}u^2\right)}}{u\mathcal{F}_\mp}.
\end{eqnarray}
Put Eq. (\ref{g9}) in Eq. (\ref{g5}), to get energy of the circular orbit
\begin{eqnarray}
E=\frac{1}{\sqrt{\mathcal{F}_\mp}}\left[1-2Mu\left(1-\tilde{\xi}u^2\right)\mp{au\sqrt{Mu\left(1-3\tilde{\xi}u^2\right)}}\right],
\end{eqnarray}
where upper and lower signs are respectively interpreted as prograde and retrograde orbits.
Angular momentum associated to the circular orbit is thus given by
\begin{eqnarray}\label{g10}
L=\frac{\mp\sqrt{M\left(1-3\tilde{\xi}u^2\right)}}{\sqrt{u\mathcal{F}_{\mp}}}\left[1+a^2u^2{\pm}{2au\left(1-\tilde{\xi}u^2\right){\sqrt{\frac{Mu}{\left(1-3\tilde{\xi}u^2\right)}}}}\right].
\end{eqnarray}
The angular velocity is computed using Eq. (\ref{t1}) and Eq. (\ref{f1})
\begin{eqnarray*}
\Omega=\frac{\dot{\phi}}{\dot{t}}=\frac{\frac{2aM}{r}\left(1-\frac{\tilde{\xi}}{r^2}\right)E+\left(1-\frac{2M}{r}\left(1-\frac{\tilde{\xi}}{r^2}\right)\right)L}{\left(r^2+a^2+\frac{2a^2M}{r}\left(1-\frac{\tilde{\xi}}{r^2}\right)\right)E-\frac{2aM}{r}\left(1-\frac{\tilde{\xi}}{r^2}\right)L},
\end{eqnarray*}
which, by using reciprocal radius, can be reduce to the form
\begin{eqnarray*}
\Omega=\frac{\left[L-2Mux\left(1-\tilde{\xi}u^2\right)\right]u^2}{\left(1+a^2u^2\right)E-2aMu^3\left(1-\tilde{\xi}u^2\right)}.
\end{eqnarray*}
This can be simplified to the form
\begin{eqnarray*}
\Omega=\frac{\mp\sqrt{Mu^3\left(1-3\tilde{\xi}u^2\right)}}{1\mp{au}\sqrt{Mu\left(1-3\tilde{\xi}u^2\right)}}.
\end{eqnarray*}
\begin{figure}[H]
    \includegraphics[width=.48\linewidth]{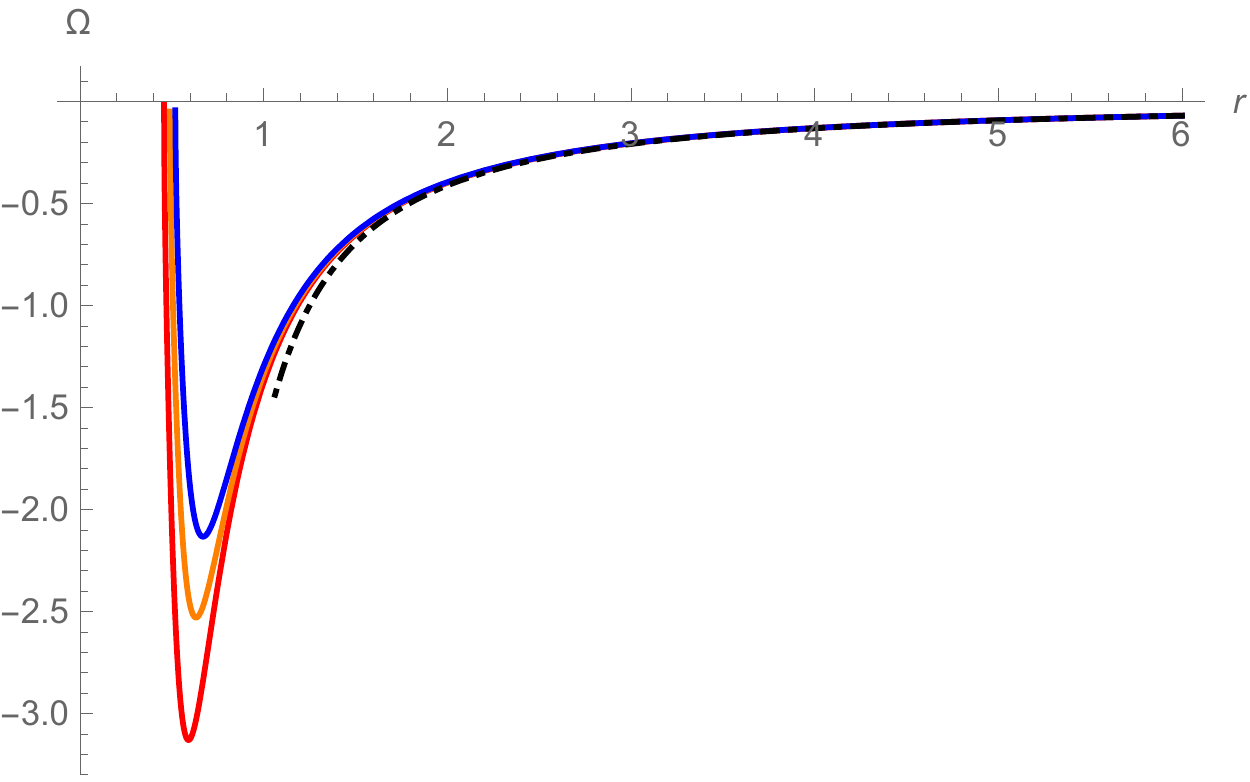}
    \includegraphics[width=.48\linewidth]{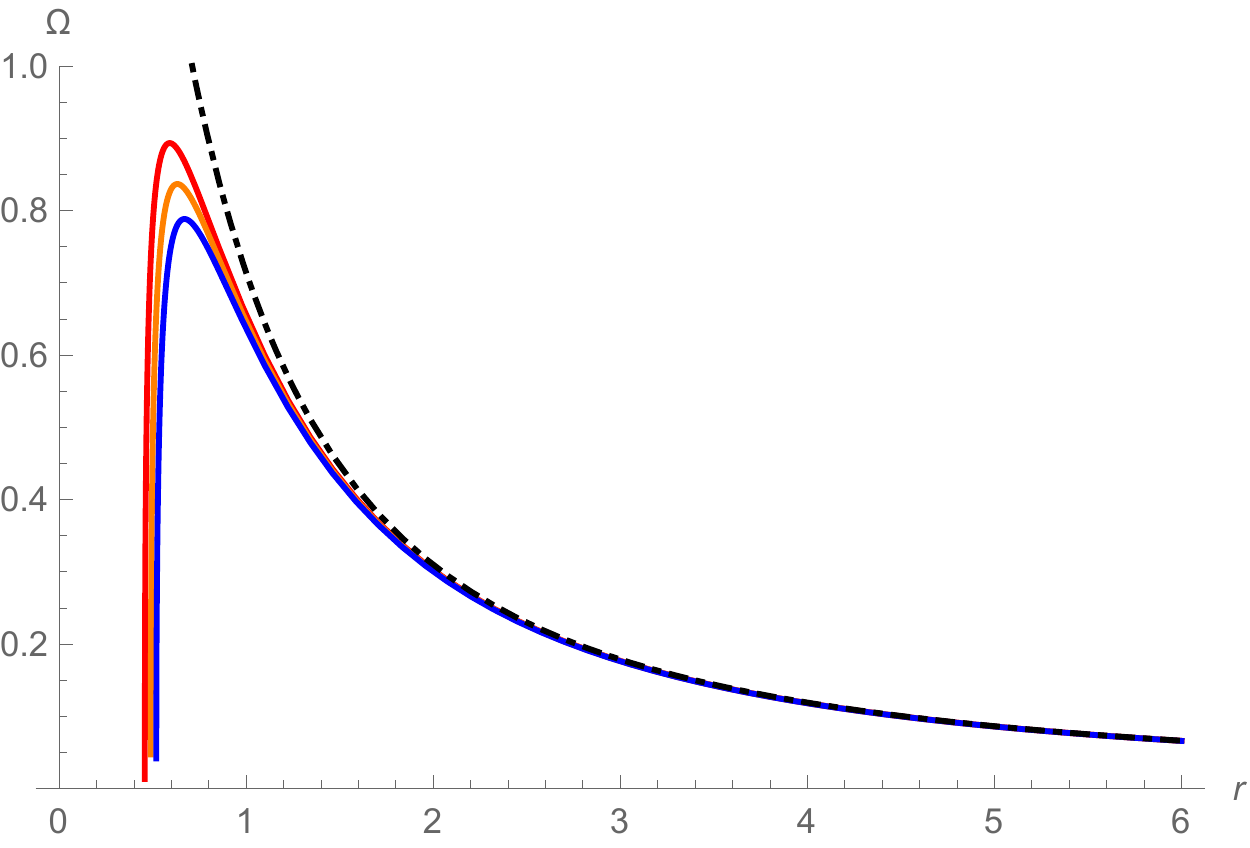}
    \caption{Angular velocity of prograde (left panel) and retrograde (right panel) motion of particles orbiting in the equatorial plane of rotating black hole in asymptotically safe gravity, for the values of the parameters as: M=1, $\tilde{\xi}=0.07$ (Red), $0.08$ (Orange), $0.09$ (Blue) and $a=0.4$. Black (DotDashed) line shows $\tilde{\xi}=0$ i.e Kerr case.}
    \label{Fig6}
    \end{figure}
Thus the angular velocity in terms of $r$ can be written, by using $r=\frac{1}{u}$, as
\begin{eqnarray*}
\Omega=\frac{\mp\sqrt{M\left(r-\frac{3\tilde{\xi}}{r}\right)}}{r^2\mp{a}\sqrt{M\left(r-\frac{3\tilde{\xi}}{r}\right)}}.
\end{eqnarray*}
The graphical representation of angular velocity of particles is shown in Figure (\ref{Fig6}). The value of $\Omega$, for corotating (prograde) motion, first decreases but it increases with the increase in $\tilde{\xi}$ and $r$. But for counter rotating (retrograde) motion, the particle's angular velocity declines when $\tilde{\xi}$ is increased.\\
The time period is given by
\begin{equation}
T=\frac{2\pi}{\Omega}=2\pi\frac{r^2\mp{a}\sqrt{M\left(r-\frac{3\tilde{\xi}}{r}\right)}}{{\mp\sqrt{M\left(r-\frac{3\tilde{\xi}}{r}\right)}}}.
\end{equation}
\subsection{Effective Potential}
To check the stability (or instability) of circular orbit of particles around the rotating black hole in asymptotically safe gravity in IR regime, the effective potential is determined. Thus the equation governing the effective potential of circular orbits, both for photons and time-like particles, is given by \cite{hartle}
\begin{eqnarray*}
\frac{E^2-1}{2}=\frac{\dot{r}^2}{2}+V_\text{eff},
\end{eqnarray*}
where effective potential is represented by $V_\text{eff}$. The extreme value $r=r_o$ of the effective potential is the solution of the equation
\begin{eqnarray*}
\frac{dV_\text{eff}}{dr}|_{r=r_o}=0.
\end{eqnarray*}
There must be present a minimum at the second derivative of effective potential i.e $\frac{d^2V_{eff}}{dr^2}>{0}$ which gives stable circular orbits along with the condition that at circular orbit $r=r_o$ the particles initial velocity must vanish i.e $\dot{r}=0$. Following is the discussion on the effective potential of null and time-like geodesics.
\subsubsection{For Null Geodesics}
For $L=aE$, the null geodesics is governed by radial equation $\dot{r}=\pm{E}$, so the case sufficient to consider here is when $L\neq{aE}$. In this case, the effective potential is thus given by
\begin{eqnarray*}
V_\text{eff}=\frac{1}{2r^3}\Big[-2M\Big(1-\frac{\tilde{\xi}}{r^2}\Big)\Big(aE-L\Big)^2+\Big(L^2-a^2E^2\Big)r-r^3\Big].
\end{eqnarray*}
\begin{figure}[H]
    \includegraphics[width=0.45\linewidth]{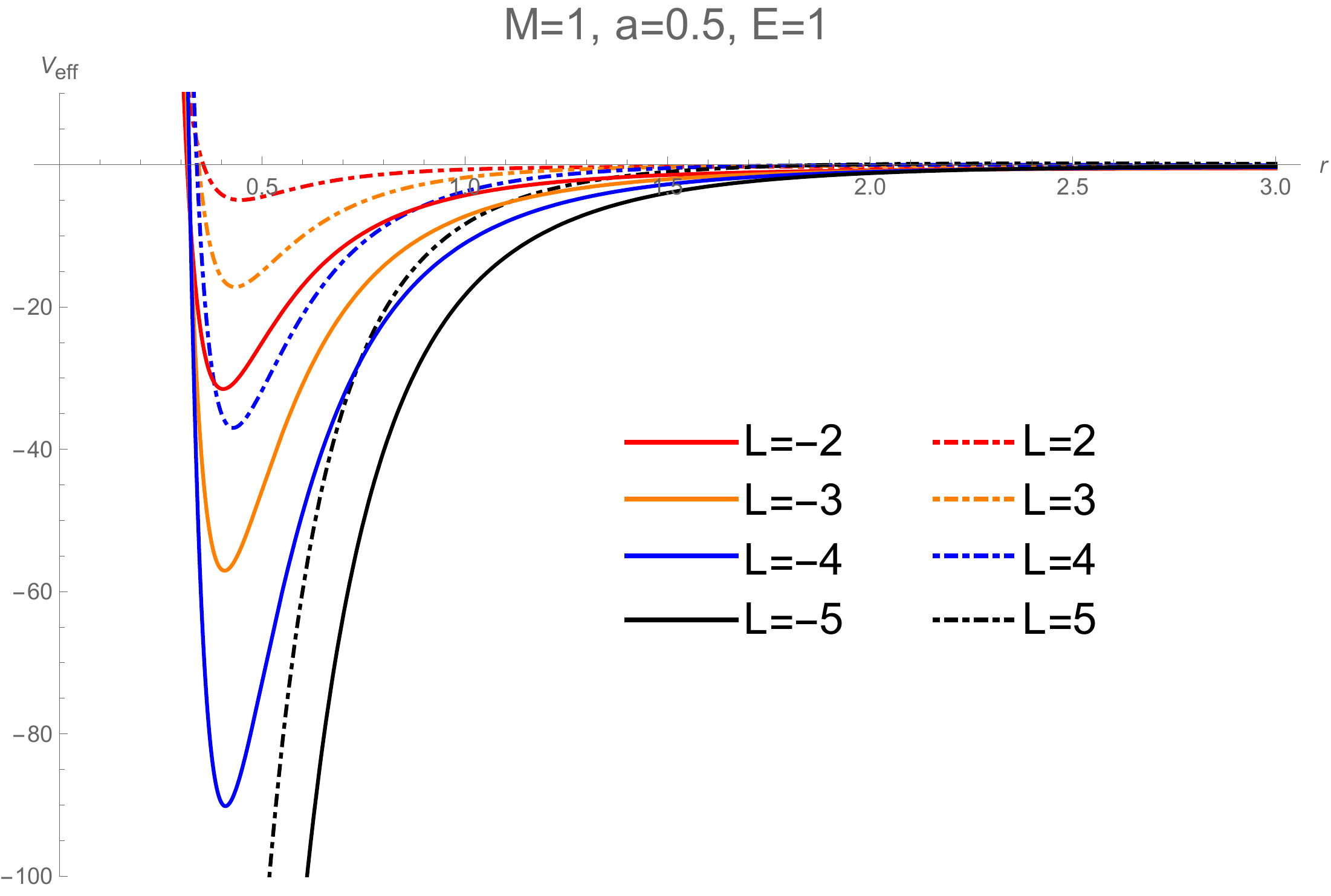}
    \includegraphics[width=0.45\linewidth]{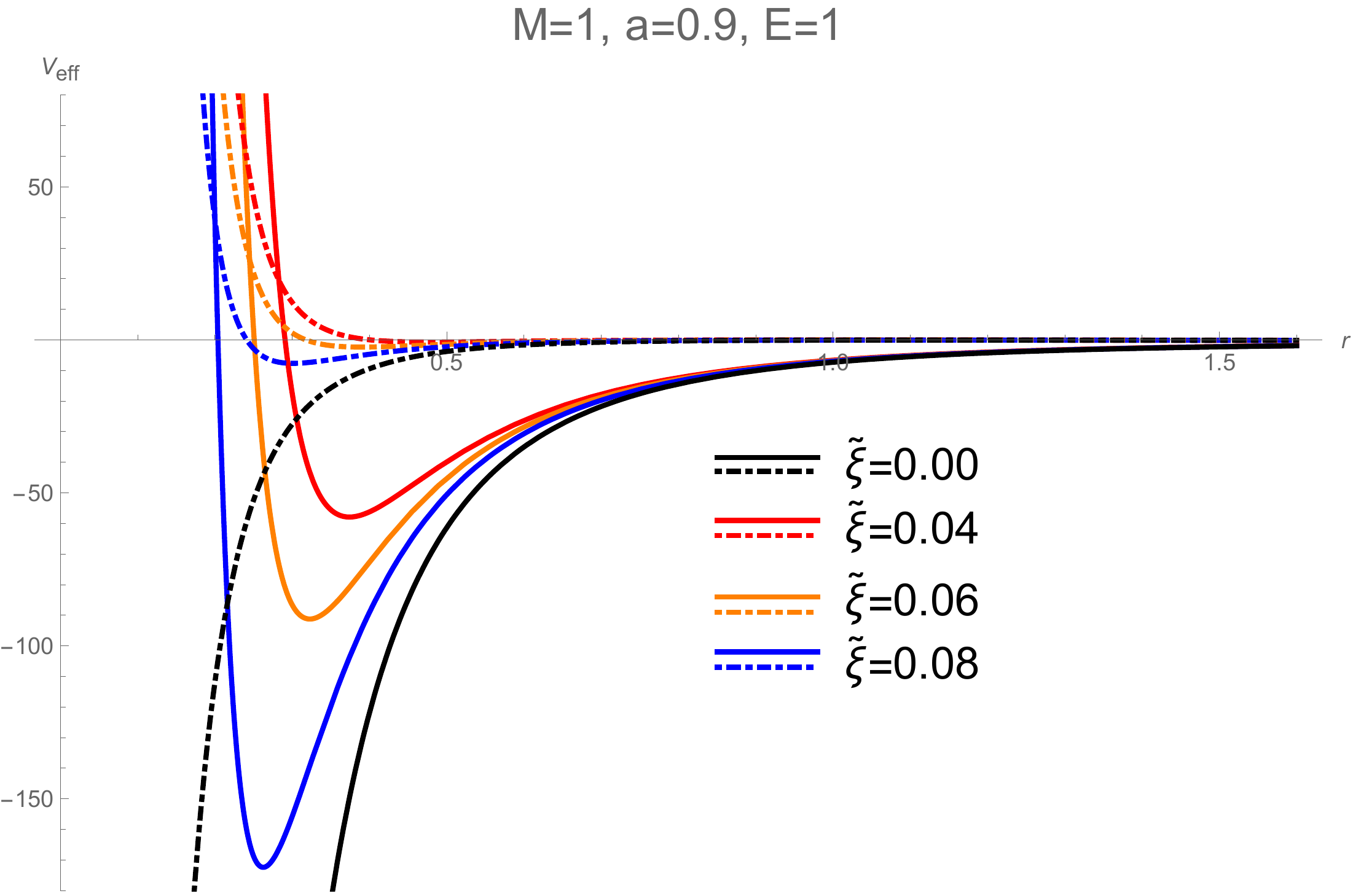}
    \caption{\scriptsize{Plots showing behaviour of effective potential, for null geodesics, with respect to $r$. Here co-rotating (counter-rotating) particles are shown by solid (dot-dashed) lines.
   }}\label{Fignl}
\end{figure}
This effective potential is graphically presented in Figure (\ref{Fignl}). As one can easily see the presence of minimum values in these plots which corresponds to the existence of stable points. Also note that the behavior of $V_\text{eff}$ for both co-rotating and counter-rotating particles is quite different from the Kerr case.
Namely, the effective potential for Kerr black hole approaches negative infinity when $r \rightarrow 0$, while the effective potential
for the rotating black hole with running gravitational coupling approaches positive infinity when $r \rightarrow 0$. This comes
as a result of sign change for the leading order in the potential for small $r$, which comes as a consequence of introducing the asymptotic correction
parameter $\tilde{\xi}$. However, it should be stressed that at very small distances IR will no longer be valid, and the proper description
should now be given using the UV limit of asymptotically safe gravity.
\subsubsection{For Time-like Geodesics}
By the use of Eq. (\ref{g1}), the effective potential for the time-like geodesics, both when $L=aE$ and $L\neq{aE}$, is computed respectively as
\begin{eqnarray*}
V_\text{eff}=\frac{a^2}{2r^2}-\frac{M}{r}\Big(1-\frac{\tilde{\xi}}{r^2}\Big)
\end{eqnarray*}
and
\begin{eqnarray*}
V_\text{eff}=\frac{-M}{r^3}\Big(1-\frac{\tilde{\xi}}{r^2}\Big)\Big(aE-L\Big)^2+\frac{L^2-a^2\Big(E^2-1\Big)}{2r^2}-\frac{M}{r}\Big(1-\frac{\tilde{\xi}}{r^2}\Big).
\end{eqnarray*}
\begin{figure}[H]
\centering
    \includegraphics[width=0.45\linewidth]{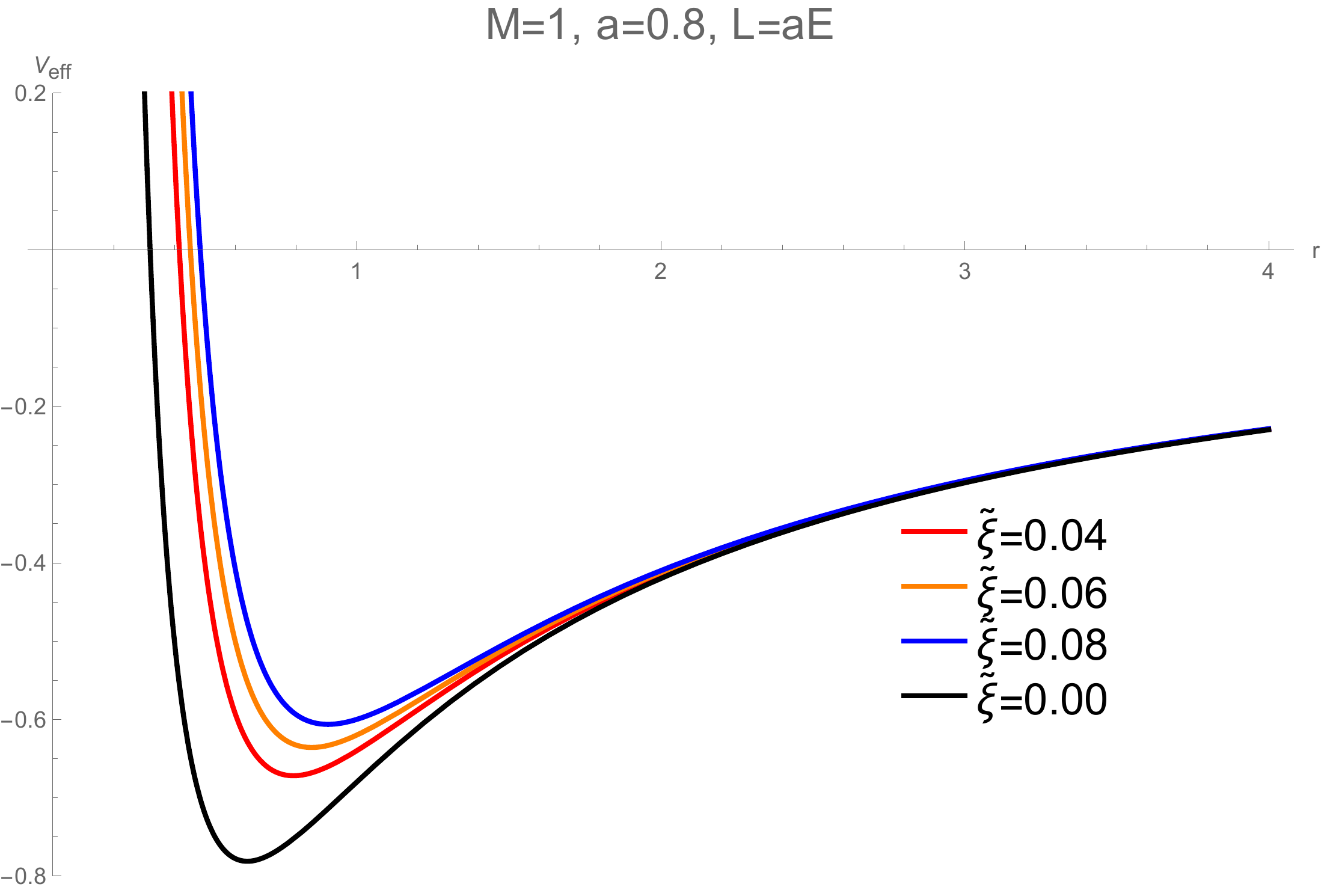}
    \caption{\scriptsize{Behaviour of effective potential for time-like geodesics, with respect to $r$, when $L=aE$. Black line shows $\tilde{\xi}=0$ i.e Kerr case.
   }}\label{Figtl1}
\end{figure}
\begin{figure}[H]
    \includegraphics[width=0.45\linewidth]{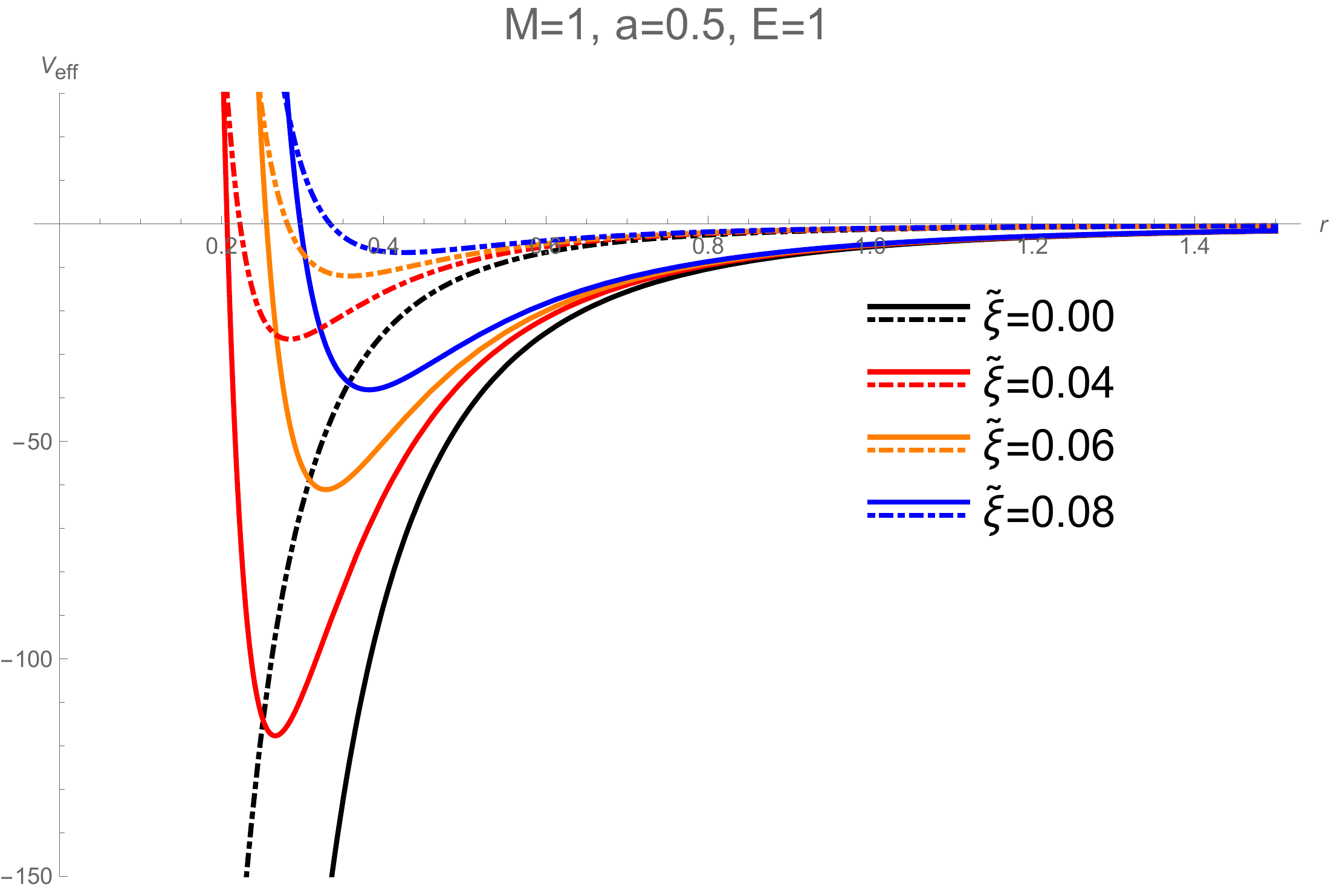}
      \includegraphics[width=0.45\linewidth]{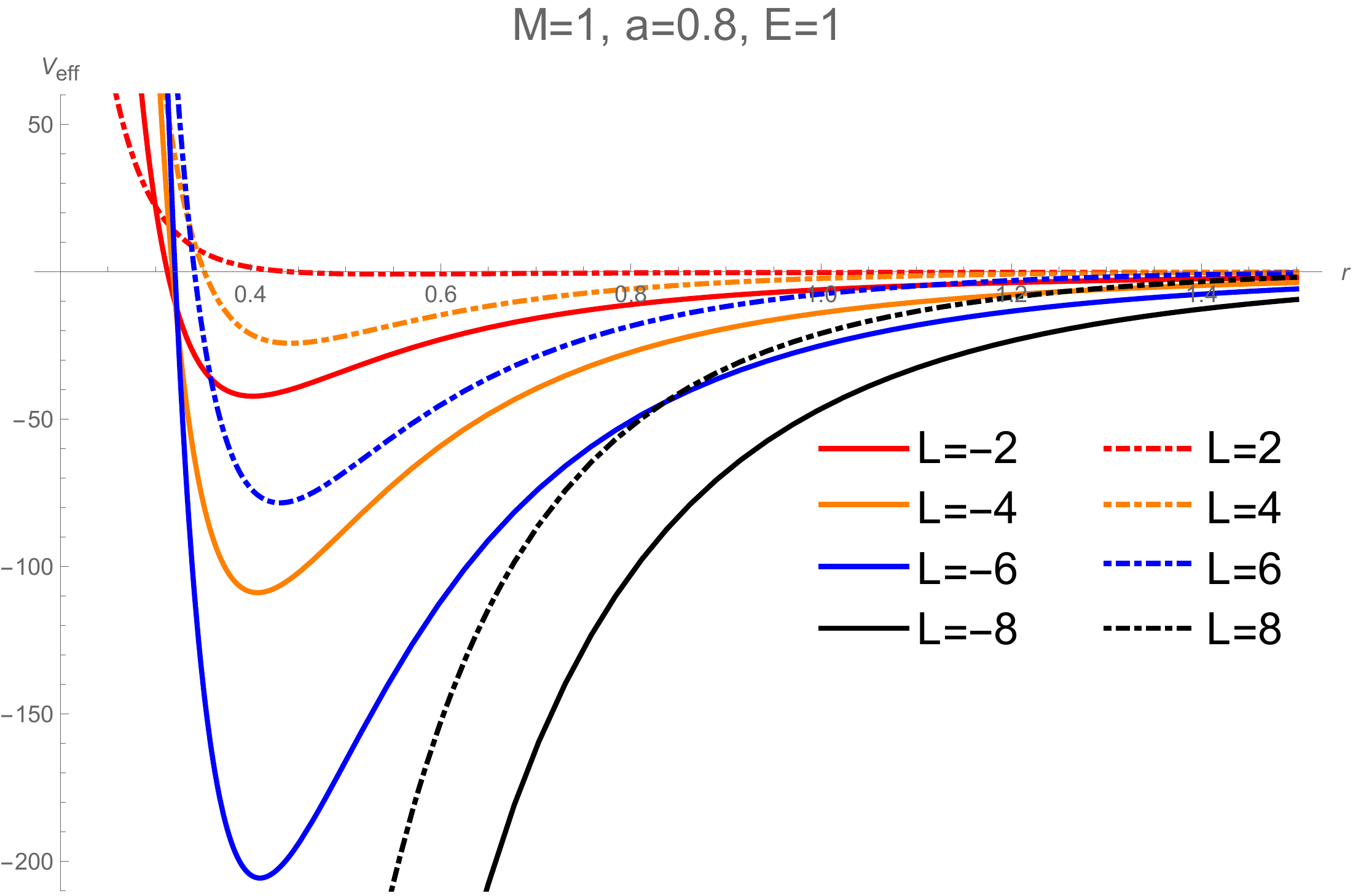}
    \caption{\scriptsize{Behaviour of effective potential for time-like geodesics, with respect to $r$, when $L\neq{aE}$. Solid lines show counter-rotating while dot-dashed lines show co-rotating particles. Black (solid/dot-dashed) line shows $\tilde{\xi}=0$ i.e Kerr case. In left panel $L$ is kept fixed ($L=-2$ and $2$ for counter-rotating and co-rotating particles) while in right panel $\tilde{\xi}=0.09$ for different values of L
   }}\label{Figtl2}
\end{figure}
For time like geodesics, when $L=aE$, the affect of $\tilde{\xi}$ on effective potential is shown in Figure (\ref{Figtl1}). These plots show existence of stable points for different values of $\tilde{\xi}$. Figure (\ref{Figtl2}) shows variation of $\tilde{\xi}$ and angular momentum $L$ in effective potential. Here it is noted that for co-rotating motion of the particle the depth of the potential well increases with increase in $L$ while for counter rotating motion, it decreases for increase in $L$. These graphs also show the existence of stable points.
As in the case of null geodesics, discussed previously, Kerr and the IR asymptotic safe solution lead to different qualitative features
of the effective potential when $r\rightarrow 0$.
\section{Penrose Process}
\begin{figure}
    \centering
    \includegraphics[width=0.46\textwidth]{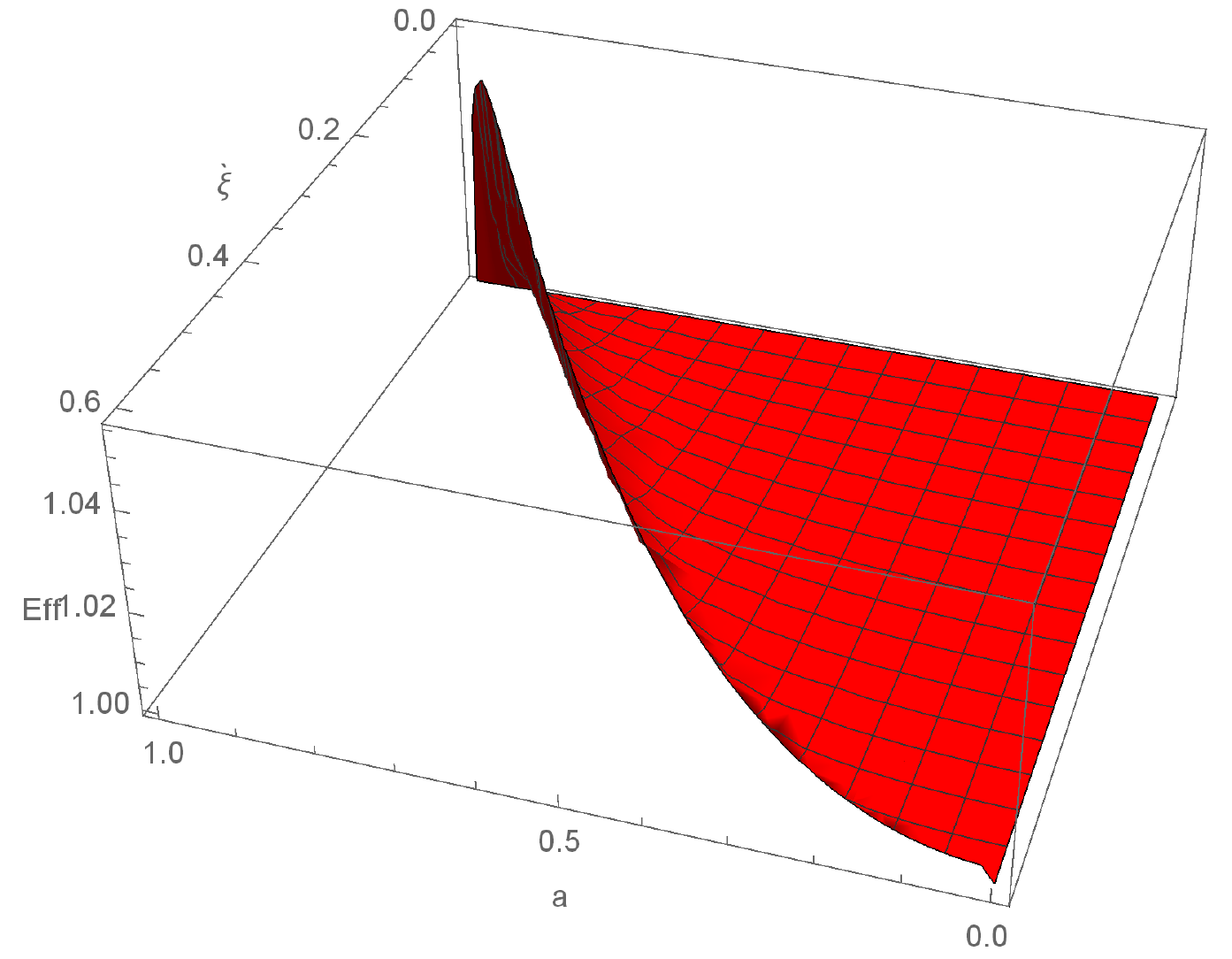}
    \caption{\scriptsize{Ratio between the efficiency of Penrose process in the IR limit of quantum corrected gravitational coupling and general relativity, as a function
    of $a$ and $\tilde{\xi}$, with $M=1$.
   }}\label{ratio}
\end{figure}

\begin{figure}
    \centering
    \includegraphics[width=0.46\textwidth]{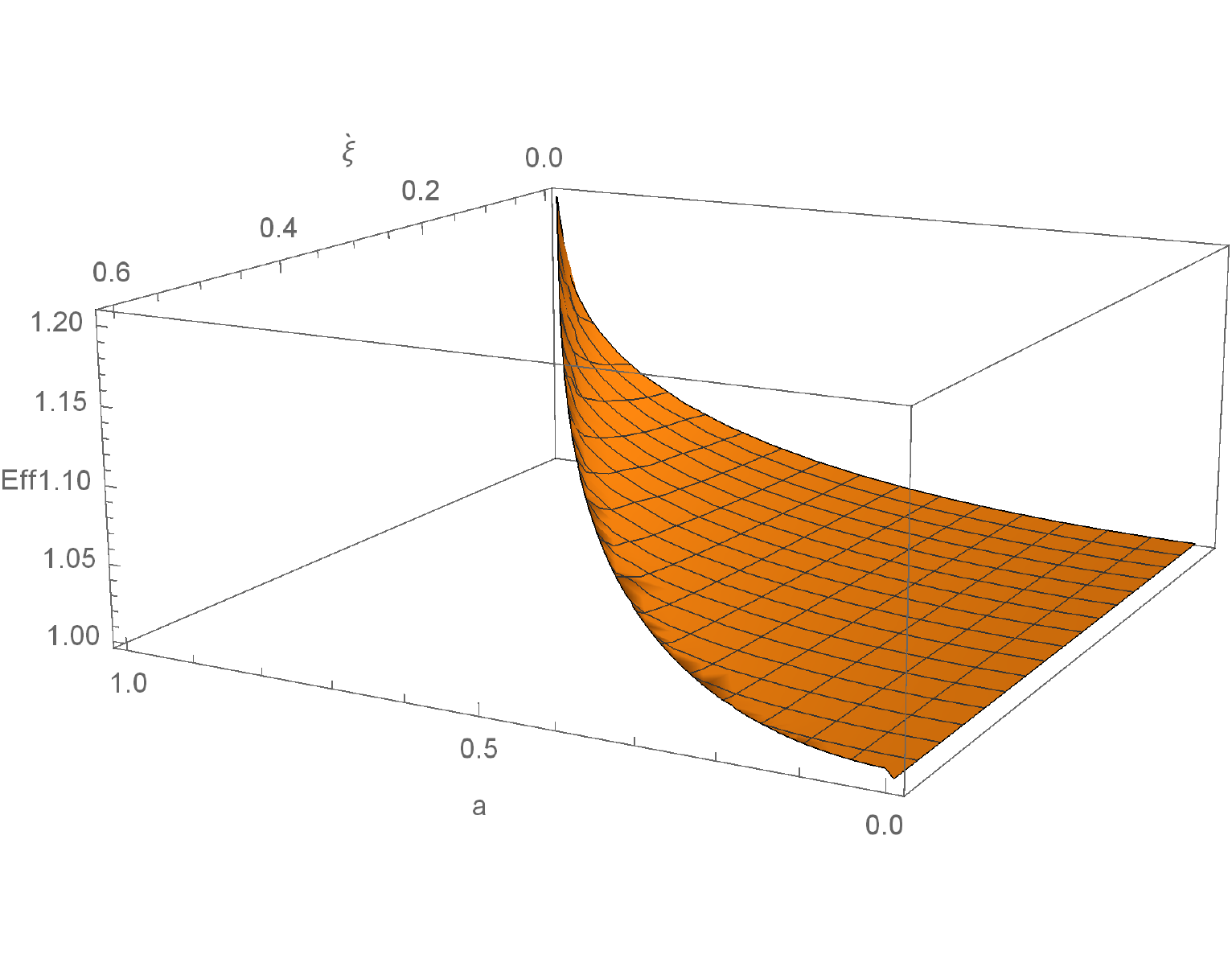}
    \caption{\scriptsize{Efficiency of Penrose process in the IR limit of quantum corrected gravitational coupling as a function
    of $a$ and $\tilde{\xi}$, with $M=1$.
   }}\label{effi}
\end{figure}

We now analyze the Penrose process, taking into account the corrections to gravitational coupling.
Let us first note, that the modifications coming from the quantum effects leading to a r-dependent gravitational coupling were
previously considered in the context of the Penrose process in \cite{RT}, where the running coupling of the following form was considered
\begin{equation}
G(r)=\frac{G_{0}r^{2}}{r^2 +\omega G_{0}} ,
\end{equation}
where $G_{0}$ is the classical Newton constant, and $\omega$ a positive constant. In \cite{RT} authors studied the functional dependencies
of tangential and dragging velocities in the Penrose process and concluded that there exists a lowest possible mass for the Penrose mechanism when
such running gravitational coupling is considered. In our analysis this is related to the mass corresponding to $\tilde{\xi}_{c}$ which
enters in equation (\ref{critical}). In this work we will perform a similar analysis of the Penrose process in the context of varying gravitational coupling, and further
extend it by investigating the efficiency of Penrose process in this setting and comparing it with the classical limit.
\\ \\
As discussed in the introduction, utilizing the fact that energy in the ergoregion can be negative, it is under suitable conditions possible to extract energy from the rotating black hole. In this discussion we concentrate on the scenario where we have a massive particle entering the ergosphere, and which moves along a timelike geodesic, carrying positive energy. This particle then decays into two particles which are massless, one carrying negative energy, and the second one with positive energy. The negative energy particle then falls into event horizon, while
the particle with a positive energy eventually leaves the ergosphere and reaches the observer. If this decay happened at the turning point of the geodesic, where
$\dot{r}=0$, then from the radial equation for equatorial geodesic it follows that
\begin{align}
E=\frac{1}{r\left(r^2+a^2\right)+2a^2M\left(1-\frac{\tilde{\xi}}{r^2}\right)}\Bigg[2aM\left(1-\frac{\tilde{\xi}}{r^2}\right)L \pm \nonumber \\
\sqrt{r^2\Delta{L^2}-\left(r\left(r^2+a^2\right)+2a^2M\left(1-\frac{\tilde{\xi}}{r^2}\right)\right)\delta{r}}\Bigg],
\end{align}
and alternatively angular momentum can be expressed as
\begin{align}
L=\frac{1}{r-2M\left(1-\frac{\tilde{\xi}}{r^2}\right)}\Bigg[-2aM\left(1-\frac{\tilde{\xi}}{r^2}\right) \pm \nonumber \\
\sqrt{\Delta{r^2}E^2+\left(1-\frac{2M}{r}\left(1-\frac{\tilde{\xi}}{r^2}\right)\right)\Delta\delta{r^2}}\Bigg],
\end{align}
where the following identity was used
\begin{eqnarray}\label{I1}
\left[r^2\left(r^2+a^2\right)+2a^2Mr\left(1-\frac{\tilde{\xi}}{r^2}\right)\right]\left(1-\frac{2M}{r}\left(1-\frac{\tilde{\xi}}{r^2}\right)\right)=r^2\Delta-4a^2M^2\left(1-\frac{\tilde{\xi}}{r^2}\right)^2.
\end{eqnarray}
Now we can determine the condition under which energy and angular momentum will be negative.
In order that $E<0$, $L<0$ it follows
\begin{eqnarray}
4a^2M^2\left(1-\frac{\tilde{\xi}}{r^2}\right)^2L^2>\Delta\left[r^2L^2-\left(r\left(r^2+a^2\right)+2a^2M\left(1-\frac{\tilde{\xi}}{r^2}\right)\right)\delta{r}\right]
\end{eqnarray}
Using eq. (\ref{I1}) this can be written as
\begin{eqnarray}
\left[r\left(r^2+a^2\right)+2a^2Mr\left(1-\frac{\tilde{\xi}}{r^2}\right)\right]\left[\left(1-\frac{2M}{r}\left(1-\frac{\tilde{\xi}}{r^2}\right)\right)L^2-\Delta\delta{r}\right]<0.
\end{eqnarray}
It follows that
$E<0 \Longleftrightarrow{L<0}$ requires the condition
\begin{eqnarray}
r\leq{2M\left(1-\frac{\tilde{\xi}}{r^2}\right)+\frac{\Delta\delta{r^2}}{L^2}},
\end{eqnarray}
so we confirm that this can happen only in the ergosphere.  \\ \\
We now come back to discussion of the decay of one initial massive particle to two massless particles, carrying the energy of opposite signs. We take that initial energy is $E_{0}=1$, and
energy of two particles is $E_{1}$ and $E_{2}$ respectively.
Let then
\begin{eqnarray}
L^{(0)}=\frac{-2aM\left(1-\frac{\tilde{\xi}}{r^2}\right)+\sqrt{2Mr\left(1-\frac{\tilde{\xi}}{r^2}\right)\Delta}}{r-2M\left(1-\frac{\tilde{\xi}}{r^2}\right)}=\alpha^{(0)},
\end{eqnarray}
\begin{eqnarray}
L^{(1)}=\frac{-2aM\left(1-\frac{\tilde{\xi}}{r^2}\right)-\sqrt{r^2\Delta}}{r-2M\left(1-\frac{\tilde{\xi}}{r^2}\right)}E^{(1)}=\alpha^{(1)}E^{(1)},
\end{eqnarray}
\begin{eqnarray}
L^{(2)}=\frac{-2aM\left(1-\frac{\tilde{\xi}}{r^2}\right)+\sqrt{r^2\Delta}}{r-2M\left(1-\frac{\tilde{\xi}}{r^2}\right)}E^{(2)}=\alpha^{(2)}E^{(2)}.
\end{eqnarray}
Here $\alpha$'s are some arbitrary functions relating angular momentum and energy.
According to conservation of energy and momentum
\begin{eqnarray}
E^{(1)}+E^{(2)}=E^{(0)}=1
\end{eqnarray}
and
\begin{eqnarray}
L^{(1)}+L^{(2)}=\alpha^{(1)}E^{(1)}+\alpha^{(2)}E^{(2)}=L^{(0)}=\alpha^{(0)} .
\end{eqnarray}
Solving these equations we can obtain the energies of two created particles as
\begin{eqnarray}
E^{(1)}=\frac{1}{2}\Big[1-\sqrt{\frac{2M}{r}(1-\frac{\tilde{\xi}}{r^2})}\Big]
\end{eqnarray}
\begin{eqnarray}
E^{(2)}=\frac{1}{2}\Big[1+\sqrt{\frac{2M}{r}(1-\frac{\tilde{\xi}}{r^2})}\Big].
\end{eqnarray}
Then, if $E^{(2)}$ reaches the observer outside the black hole, and $E^{(1)}$ crosses the event horizon, the gain in energy with respect to the original particle, as measured by
the observer is
\begin{eqnarray}
\Delta{E}=\frac{1}{2}\Big[\sqrt{\frac{2M}{r}(1-\frac{\tilde{\xi}}{r^2})}-1\Big]=-E^{(1)} .
\label{gain}
\end{eqnarray}
We stress that the analysis above can simply be reduced to classical case of a constant gravitational coupling, just by taking 
$\tilde{\xi}=0$.
In order to study the maximal possible efficiency of Penrose process one should consider the case with respect to which any reasonable physical realization will lead to smaller values.
The gain in energy will be bigger if the radial distance is smaller, so we consider the extreme case of the event horizon $r=r_{H}$. For simplicity, we can use for example black hole defined by
$r_{H}=M=1$ . For such black hole from the horizon equation it follows that $\tilde{\xi}=\frac{1}{2}(1-a^{2})$. We get that the maximal efficiency of Penrose process in this case is then given by
\begin{equation}
E_{\text{ff}_{max}}=\frac{E_{0} +\Delta{E}}{E_{0}}_{max}=\frac{1}{2}[1+\sqrt{2(1-\tilde{\xi})}] < 1.207,
\label{eff}
\end{equation}
so we see that for a given black hole with the same fixed parameters $r_{H}$ and $M$ in general relativity and IR limit of asymptotically safe gravity, the efficiency of Penrose process will be smaller when the effect of running gravitational 
coupling is considered. However,
as noted earlier, in the quantum corrected case the outer event horizon tends to be located at smaller $r$ than in the standard general relativity, for all other parameters staying the same.
This fact can thus compensate the direct loss coming from the corrective term in Eq. (\ref{gain}), and can even increase the efficiency
above the one characteristic for Kerr black hole in general relativity. We should stress that, from the astrophysical perspective, $a$ and
$M$ should be considered as real independent quantities defining the black hole -- actually given as initial conditions during the collapse of matter leading
to black hole formation -- and that position of event horizon should be considered as a dependent quantity. Therefore, it is more proper to compare
rotating black holes with running Newtonian coupling and general relativity for the same values of $M$ and $a$, rather than $r_{H}$. Taking $M=1$
for simplicity, we show that -- in accord with the previous reasoning -- for the same values of $a$ the efficiency of Penrose process
will be greater when the IR corrections are introduced. This is demonstrated in Figure \ref{ratio} where we show the ratio of efficiency of Penrose process in quantum corrected
 gravity and general relativity as a function of $a$ and $\tilde{\xi}$. However, it can be seen in Figure \ref{effi} -- where we plot the efficiency
of Penrose process when IR correction is included, that the maximal possible efficiency still
basically stays confined within the region estimated in Eq. (\ref{eff}). Finishing the discussion on Penrose process let us futher stress
the comparison with the classical case. We have shown that for a black hole with given values of outer event horizon and mass the efficiency 
of Penrose process will be smaller in IR limit of quantum corrected general relativity than in the classical case. However, if the values of 
$a$ and $M$ are taken as given, and they should in fact be considered as a more physical input parameters since they are related to dynamics 
of the body from which the black hole can be formed, than the outer event horizon will typically be smaller including the IR corrections, and 
thus the efficiency can be even increased with respect to the one characterizing the classical case. 
\section{Lense-Thirring frequency }

The forms of electromagnetic equations and gravitational equations are very similar, so that the gravito-electromagnetism \cite{gravitoeletromagnetism} summarizes the weak field gravitational equations as the ``Maxwell Equations".
 As we all know, a rotating sphere with electric charge can produce magnetic field, so it is reasonable to believe that ``magnetic effect" of gravitational field can be found in spacetime with rotating massive sphere.
In 1918, Lense and Thirring theoretically proposed Lense-Thirring effect to describe the ``magnetic effect" in gravitational field \cite{LenseThirring}.
According to \cite{OmegaLT1,OmegaLT2,OmegaLT3,OmegaLT4}, the precession frequency vector of rotating black hole is given by
\begin{eqnarray}
\label{LT1}
\Omega_{LT}=\frac{1}{2}\frac{\epsilon_{ijl}}{\sqrt{-g}}\left[g_{0i,j}\left(\partial_l-\frac{g_{0l}}{g_{00}}\partial_0\right)-\frac{g_{0i}}{g_{00}}g_{00,j}\partial_l\right].
\end{eqnarray}
In this paper, from our metric, above result is rewritten as
\begin{eqnarray}
\label{LT2}
\Omega_{LT}=\frac{1}{2\sqrt{-g}}\left[\left(g_{0\phi,r}-\frac{g_{0\phi}}{g_{00}}g_{00,r}\right)\partial_\phi-\left(g_{0\phi,\theta}-\frac{g_{0\phi}}{g_{00}}g_{00,\theta}\right)\partial_r\right]
\end{eqnarray}

\begin{eqnarray}
\label{LT3}
\Omega_{LT}&=&\Omega^\theta\partial_\theta+\Omega^r\partial_r,\nonumber\\
\Omega_{LT}^2&=&g_{rr}\left(\Omega^r\right)^2+g_{\theta\theta}\left(\Omega^\theta\right)^2.
\end{eqnarray}
While in polar coordinates (where $\hat{r}$ is the unit vector of
direction $r$ and $\hat{\theta}$ is angular coordinate),
$\Omega_{LT}$ is given by
\begin{eqnarray}
\label{LT4}
{\stackrel{\rightarrow}{\Omega}}_{LT}=\sqrt{g_{rr}}\Omega^r\hat{r}+\sqrt{g_{\theta\theta}}\Omega^\theta\hat{\theta}.
\end{eqnarray}
Therefore, for our black hole spacetime
\begin{eqnarray}
\label{LT5}
\Omega^\theta&=&\frac{2aM\left(r^2-\tilde{\xi}\right)\left(a^2r-2Mr^2+r^3+2M\tilde{\xi}\right)\cos(\theta)}{r\left(r^2+a^2\cos\left(\theta\right)^2\right)^2\left(r^3-2Mr^2+2M\tilde{\xi}+a^2r\cos\left(\theta\right)^2\right)},\nonumber\\
\Omega^r&=&\frac{aM\left[r^2\left(r^2-3\tilde{\xi}\right)\sin\left(\theta\right)-a^2\left(r^2+\tilde{\xi}\right)\cos\left(\theta\right)^2\sin\left(\theta\right)\right]}{r\left(r^2+a^2\cos\left(\theta\right)^2\right)^2\left(r^3-2Mr^2+2M\tilde{\xi}+a^2r\cos\left(\theta\right)^2\right)}.
\end{eqnarray}

Therefore the magnitude of $\Omega_{LT}$ is given by
\begin{eqnarray}
\label{LT6}
\Omega_{LT}=\Omega_{\text{StrongLT}}=J\frac{\sqrt{4r^2\left(r^2-\tilde{\xi}^2\right)^2\Delta\cos\left(\theta\right)^2+\left(r^4-3\tilde{\xi}
r^2-a^2\left(r^2+\tilde{\xi}\right)\cos\left(\theta\right)^2\right)^2\sin\left(\theta\right)^2}}{r^2\left(\Delta-a^2\sin\left(\theta\right)^2\right)\Sigma^{3/2}}
\end{eqnarray}
According to \cite{OmegaLT1}, in the weak field limit (which means $r\gg M$), we expand above formula by $M$, so $\Omega_{LT}$ in weak field is:
\begin{eqnarray}
\label{LT6}
\Omega_{\text{weakLT}}&=&\frac{J}{r^2\Sigma^{5/2}}\left\{r^2\cos\left(\theta\right)^2\left[4\left(r^3-r\tilde{\xi}\right)^2+a^2\left(3r^4-8\tilde{\xi}
r^2+7\tilde{\xi}^2\right)+a^2\left(r^4-3\tilde{\xi}^2\right)\cos\left(2\theta\right)\right]\right.\nonumber\\
&&\left.+a^4\left(r^2+\tilde{\xi}\right)^2\cos\left(\theta\right)^4\sin\left(\theta\right)^2+r^4\left[\left(r^2-3\tilde{\xi}\right)^2+4a^2\tilde{\xi}\cos\left(\theta\right)^2\right]\sin\left(\theta\right)^2\right\}^{1/2}+{\cal
O}\left(M^2\right)\nonumber\\
\end{eqnarray}

\begin{figure}
    \centering
    \includegraphics[width=0.46\textwidth]{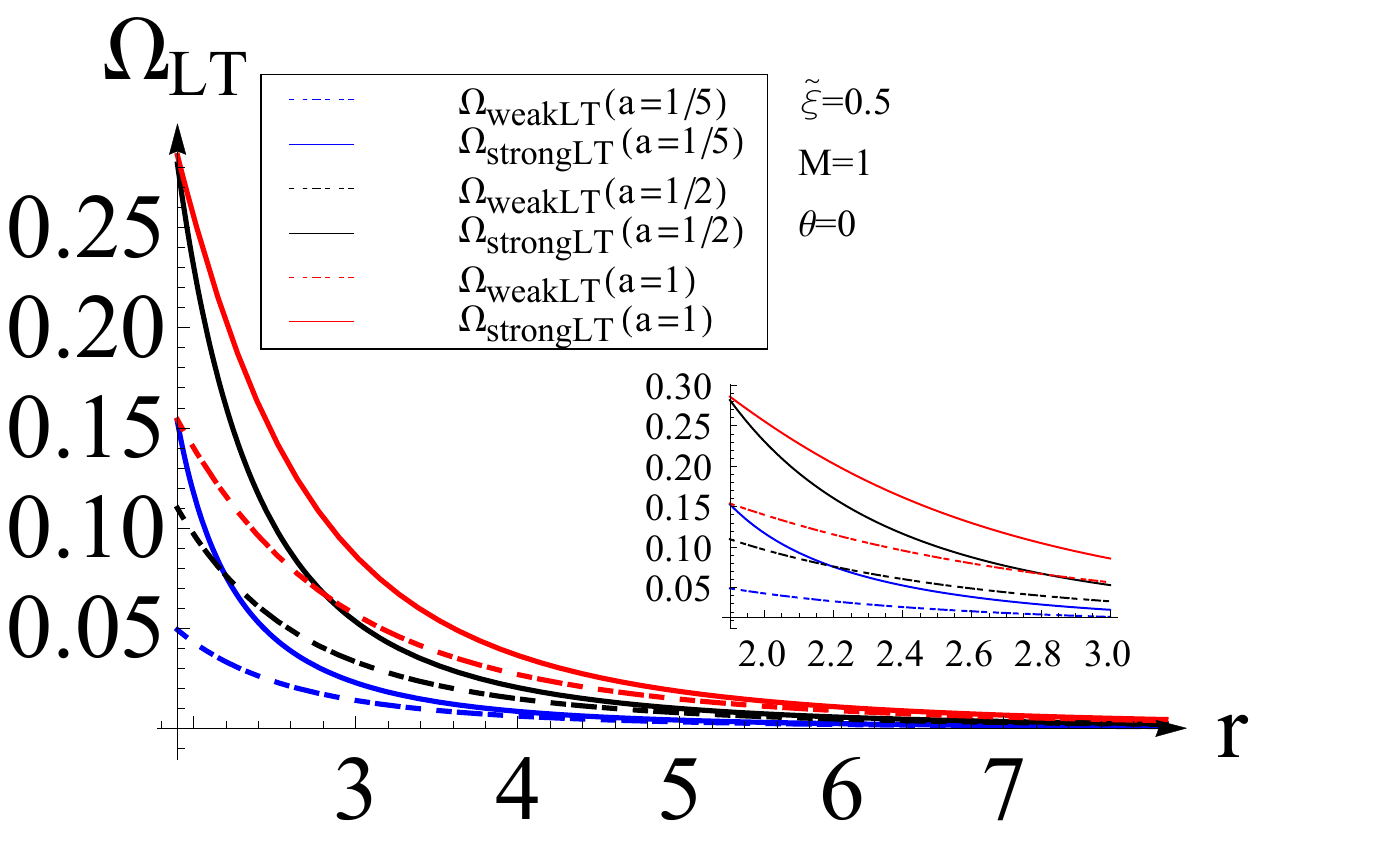}
    \caption{$\Omega_{\text{strongLT}}$ and $\Omega_{\text{weakLT}}$ as functions of r for $a=1/5,1/2,1$, where $\tilde{\xi}=0.5$, $M=1$, $\theta=0$}\label{Fig3}
\label{LTfig1}
\end{figure}

\begin{figure}
    \centering
    \includegraphics[width=0.46\textwidth]{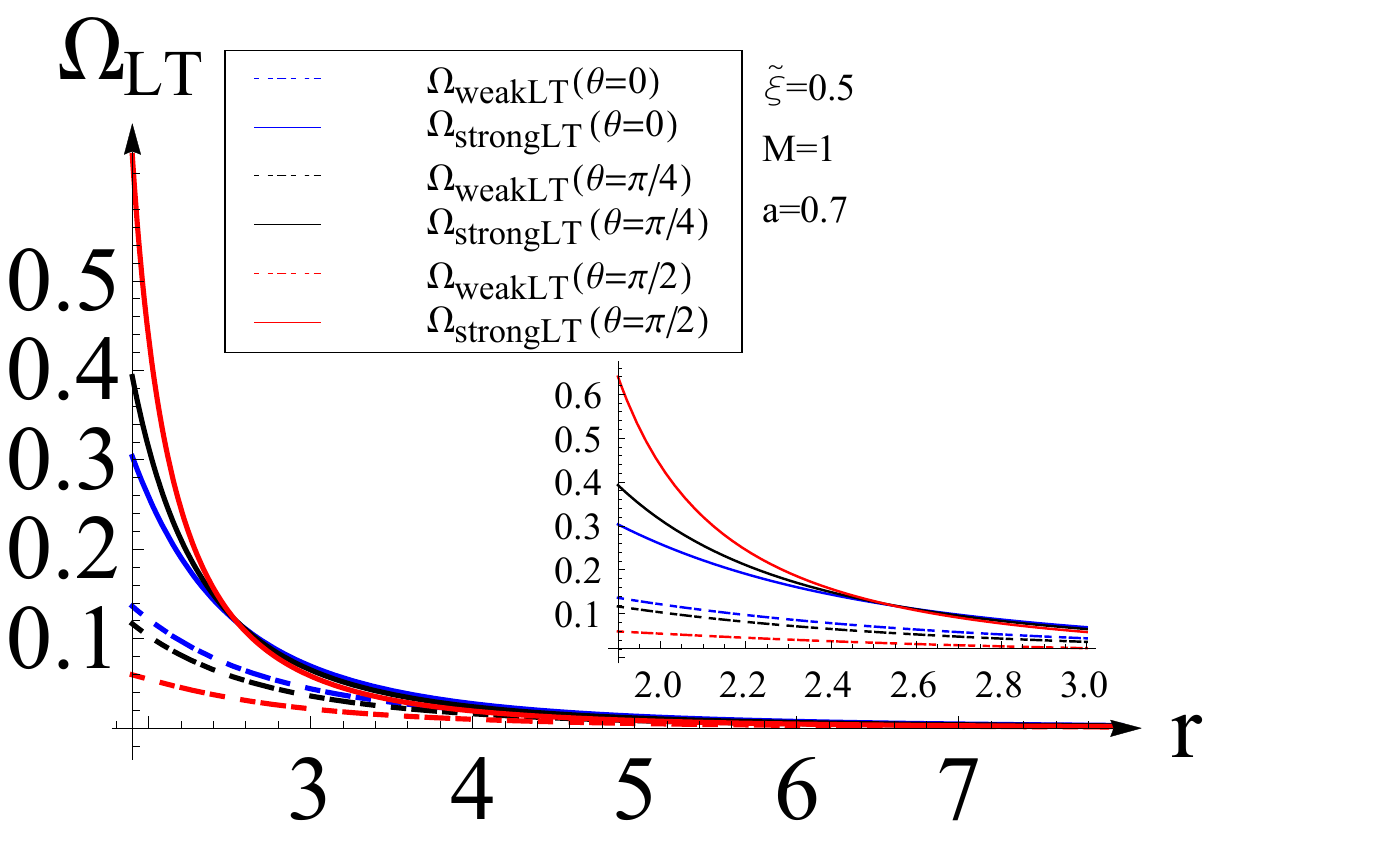}
    \caption{$\Omega_{\text{strongLT}}$ and $\Omega_{\text{weakLT}}$ as functions of r for $\theta=\pi/2,\pi/4,0$, where $\tilde{\xi}=0.5$, $M=1$, $a=0.7$}\label{Fig4}
\label{LTfig2}
\end{figure}

\begin{figure}
    \centering
    \includegraphics[width=0.46\textwidth]{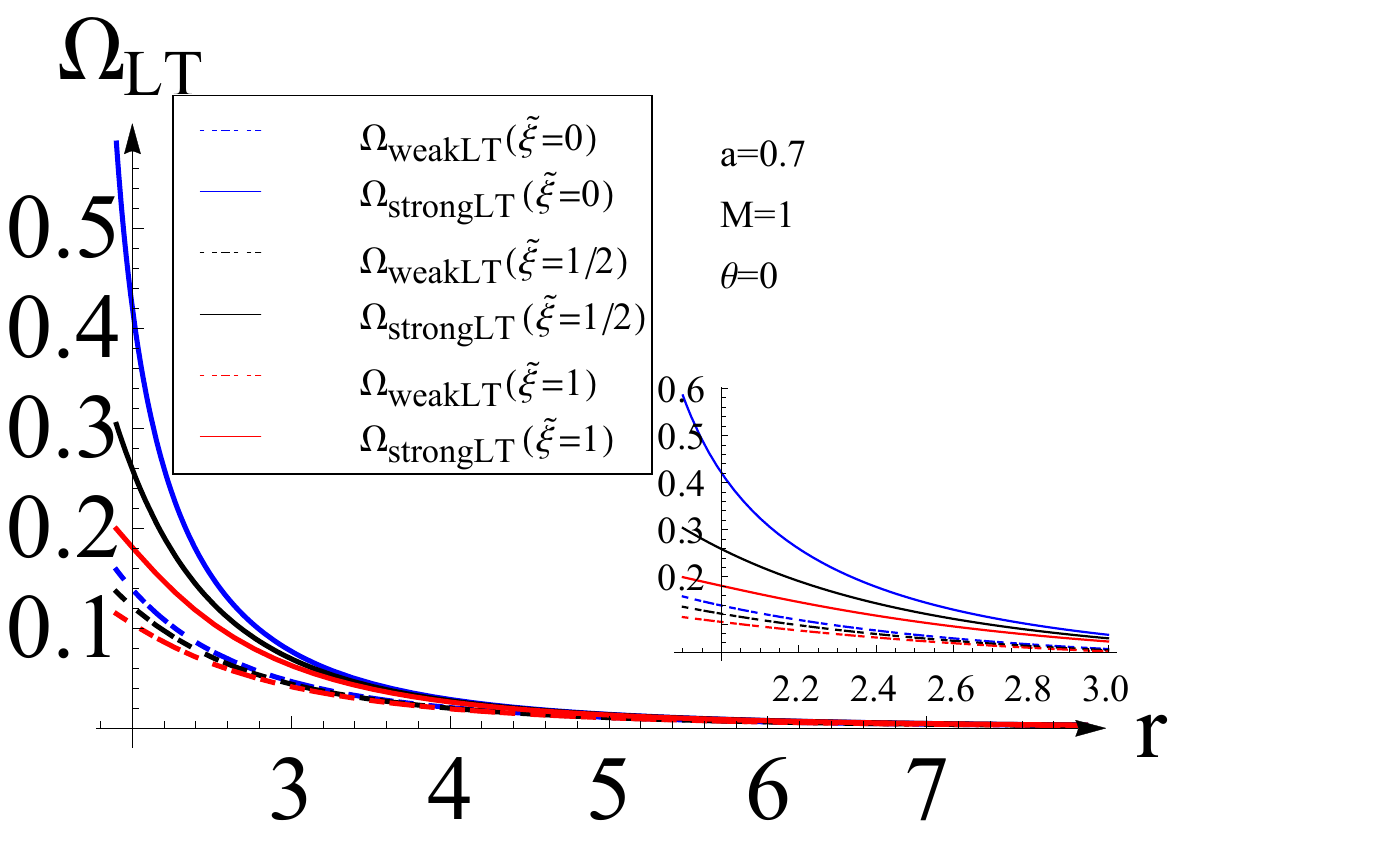}
    \caption{$\Omega_{\text{strongLT}}$ and $\Omega_{\text{weakLT}}$ as functions of r for $\tilde{\xi}=0,1/2,1$, where $a=0.7$, $M=1$, $\theta=0$}\label{Fig4}
\label{LTfig3}
\end{figure}

We show the $\Omega_{LT}=\Omega_{LT}(r)$ with various parameters in Figure (\ref{LTfig1}), Figure (\ref{LTfig2}) and Figure (\ref{LTfig3}).

The Figure (\ref{LTfig1}) shows that Lense-Thirring effect is significantly increased as $a$ is increased because it is a rotating effect.
On the other hand, according to the Figure (\ref{LTfig2}), it is interesting that the effect of $\Omega_{\text{strongLT}}$ is more outstanding at equator $\theta=\pi/2$ than the pole $\theta=0$, but the effect of $\Omega_{\text{weakLT}}$ is more outstanding at the pole
$\theta=0$ than equator $\theta=\pi/2$.
Finally, from Figure (\ref{LTfig3}), we find that Lense-Thirring effect in our rotating black hole spacetime is weaker than Kerr spacetime as $\tilde{\xi}$ increases, so it means that it is more difficult to measure this effect in our metric, but we can compare the results of experiment
to determine the value of $\tilde{\xi}$, and rotating spacetime won't be Kerr spacetime if $\tilde{\xi}\not=0$.

\section{Discussion}
In this work we have discussed in detail the properties of a rotating black hole solution considering the variation of gravitational coupling. This new metric is characterized by three parameters: rotational parameter $a$, mass of black hole $M$ and constant parameter $\tilde{\xi}$, describing the quantum corrections. We have first discussed the consequences of running gravitational coupling on the structure of event and Killing horizons. While doing so it is noted that ergosphere is significantly increased when $\tilde{\xi}$ increases, which also depicts that the region of the black hole from which energy can be extracted, through Penrose process, is bigger as compared to standard GR. It should be kept in mind here that to avoid the whole system to get strongly coupled, Cai and Easson \cite{P} considered the value of coupling parameter $\tilde{\xi}$ to be less than unity, which limitized the practical significance of our result. Further, particle's motion is investigated by studying geodesics for both null and time-like particles. For the case $L=aE$, the equations for outgoing photon trajectory are numerically solved. By plotting these results, it is observed that,for small $\tilde{\xi}$, the photon trajectory with respect to both time $t$ and angle $\phi$ shows no high deviation from its GR counterpart. The presence of a bifurcation point is also numerically analyzed, which leads us to existence of naked singularity. Expressions of energy and angular momentum for time-like geodesics, depending on $r$, are derived. Angular velocity $\Omega$ is computed and it is observed that for prograde motion $\Omega$ increases when $\tilde{\xi}$ is increased but for retrograde motion it decreases with $\tilde{\xi}$. A detailed discussion is made on effective potential also. It is graphically shown that the shape of potential barrier is changed for different values of $\tilde{\xi}$. The extraction of energy is discussed by taking into account the Penrose process. We have demonstrated that for the same values of $a$ the efficiency of Penrose process
will be greater in IR limit of quantum corrected gravity, while the maximum possible efficiency of Penrose process will not be significantly changed. In the end, another effect called Lense-Thirring effect is also explored. It is noted that this effect, being a rotational effect, depends on the rotational parameter $a$. With the increase in the value of $a$ Lense-Thirring effect shows notable change when plotted.
\section{Appendix}
To compute a rotating black hole in asymptotically safe gravity, the method given in \cite{A} is subsequently adapted.
We first consider a static and spherically symmetric metric given by \cite{P}
\begin{eqnarray}\label{1}
ds^2 &=&-A(r)dt^2+\frac{dr^2}{B(r)}+h(r)\left(r^2d\theta^2+r^2\sin^2\theta{d\phi^2}\right),
\end{eqnarray}
where
\begin{eqnarray}\nonumber
A(r)=B(r)\simeq1-\frac{2M}{r}\left(1-\frac{\tilde{\xi}}{r^2}\right)~~\text{and}~~h(r)=r^2.
\end{eqnarray}
The algorithm, for metric's computation, begins by transforming Boyer-Lindquist (BL) coordinates $(t,r,\theta,\phi)$ to Eddington-Finkelstein (EF) coordinates $(u,r,\theta,\phi)$. So with the use of the following coordinate transformation, in Eq. (\ref{1}),
\begin{eqnarray*}
dt&=&du+\frac{dr}{\sqrt{AB}},
\end{eqnarray*}
a line element of the form
\begin{eqnarray*}
ds^{2}&=&-Adu^2-2\sqrt{\frac{A}{B}}dudr+h(r)d\Omega^2,
\end{eqnarray*}
is obtained. This metric in terms of null tetrads is written as
\begin{eqnarray}\label{d}
g^{\mu{\nu}}=-l^{\mu}n^{\nu}-l^{\nu}n^{\mu}+m^{\mu}\bar{m}^{\nu}+m^{\nu}\bar{m}^{\mu},
\end{eqnarray}
where null tetrads are
\begin{eqnarray*}
l^{\mu}&=&\delta^{\mu}_{r},\\
n^{\mu}&=&\sqrt\frac{B}{A}\delta^{\mu}_{u}-\frac{B}{2}\delta^{\mu}_{r},\\
m^\mu&=&\frac{1}{\sqrt{2h}}(\delta^{\mu}_{\theta}+\frac{\dot{\iota}}{\sin\theta}\delta^{\mu}_{\phi}),\\
\bar{m^\mu}&=&\frac{1}{\sqrt{2h}}(\delta^{\mu}_{\theta}-\frac{\dot{\iota}}{\sin\theta}\delta^{\mu}_{\phi}).
\end{eqnarray*}
The first two null tetrads, $l$ and $m$, are real vectors while $m$ is complex and $\bar{m}$ is conjugate of vector $m$. These vectors are orthogonal, isotropic and normalized i.e they satisfy the following conditions
\begin{eqnarray*}\nonumber
l^{\mu}l_{\mu}=n^{\mu}n_{\mu}=m^{\mu}m_{\mu}=\bar{m}^{\mu}\bar{m}_{\mu}=0,\\
l^{\mu}m_{\mu}=l^{\mu}\bar{m}_{\mu}=n^{\mu}m_{\mu}=n^{\mu}\bar{m}_{\mu}=0,\\
l^{\mu}n_{\mu}=-m^{\mu}\bar{m}_{\mu}=1.
\end{eqnarray*}
Introducing complex coordinate transformations
\begin{eqnarray*}
u'\rightarrow{u-\dot{\iota}{a}\cos\theta},\\
r'\rightarrow{r+\dot{\iota}{a}\cos\theta},
\end{eqnarray*}
where $a$ is the rotational parameter. It is also assumed that due to these transformations the functions $A(r)$, $B(r)$ and $h(r)$ shift to $F=F(r,a,\theta)$, $G=G(r,a,\theta)$ and $\Sigma=\Sigma(r,a,\theta)$ respectively.\\
This leads to new null tetrads (dropping primes) as
\begin{eqnarray}\nonumber
l^{\mu}&=&\delta^{\mu}_{r},\\\nonumber
n^{\mu}&=&\sqrt\frac{G}{F}\delta^{\mu}_{u}-\frac{G}{2}\delta^{\mu}_{r},\\\label{e11}
m^{\mu}&=&\frac{1}{\sqrt{2\Sigma}}[(\delta^{\mu}_{u}-\delta^{\mu}_{r})\dot{\iota}{a}\sin\theta+\delta^{\mu}_{\theta}+\frac{\dot{\iota}}{\sin\theta}\delta^{\mu}_{\phi}],\\\nonumber
\bar{m^{\mu}}&=&\frac{1}{\sqrt{2\Sigma}}[-(\delta^{\mu}_{u}-\delta^{\mu}_{r})\dot{\iota}{a}\sin\theta+\delta^{\mu}_{\theta}-\frac{\dot{\iota}}{\sin\theta}\delta^{\mu}_{\phi}].
\end{eqnarray}
With the help of Eq. (\ref{d}) and Eq. (\ref{e11}), contravariant components of new metric are computed as
\begin{eqnarray*}
g^{uu}&=&\frac{a^{2}\sin^{2}\theta}{\Sigma},\   \ g^{u\phi}=\frac{a}{\Sigma}, \   \ g^{ur}=-\sqrt{\frac{G}{F}}-\frac{a^{2}\sin{2}\theta}{\Sigma},\\
g^{rr}&=&G+\frac{a^{2}\sin^{2}\theta}{\Sigma},\   \ g^{r\phi}=-\frac{a}{\Sigma},\   \ g^{\theta\theta}=\frac{1}{\Sigma},\\
g^{\phi\phi}&=&\frac{1}{\Sigma\sin^2\theta}.
\end{eqnarray*}
Using the above contravariant components, the non-zero covariant components are
\begin{eqnarray}\nonumber
g_{uu}&=&-F,\   \ g_{ur}=-\sqrt{\frac{F}{G}},\   \ g_{u\phi}=a\sin^2\theta\left(F-\sqrt{\frac{F}{G}}\right),\\
g_{r\phi}&=&a\sqrt{\frac{F}{G}}\sin^2\theta,\  \ g_{\theta\theta}=\Sigma,\ \ g_{\phi\phi}=\sin^2\theta\left[\Sigma-a^2\left(F-2\sqrt{\frac{F}{G}}\right)\sin^2\theta\right].\nonumber
\end{eqnarray}
So new metric is
\begin{eqnarray*}
ds^2&=&-Fdu^2-2\sqrt{\frac{F}{G}}dudr+2a\sin^2\theta\left(F-\sqrt{\frac{F}{G}}\right)du{d\phi}+2a\sin^2\sqrt{\frac{F}{G}}drd\phi+\Sigma d\theta^2\\ &+&\sin^2\theta\left[\Sigma-a^2\left(F-2\sqrt{\frac{F}{G}}\right)\sin^2\theta\right]d\phi^2.
\end{eqnarray*}
Finally, the EF coordinates are transformed back to BL coordinates. For this purpose the following transformation is being used
\begin{eqnarray}\nonumber
du&=&dt+\lambda(r)dr,\\\nonumber
d\phi&=&d\phi'+\chi(r)dr,
\end{eqnarray}
where
\begin{eqnarray*}
{\lambda(r)}=\frac{-a^2-k(r)}{B(r)h(r)+a^2},\   \ {\chi(r)}=\frac{-a}{B(r)h(r)+a^2},\     \ k(r)={\sqrt\frac{B(r)}{A(r)}}h(r),
\end{eqnarray*}
with
\begin{eqnarray}
F&=&\frac{B(r)h(r)+a^2\cos^2\theta}{\left(k(r)+a^2\cos^2\theta\right)^2}\Sigma\label{8}
\end{eqnarray}
and
\begin{eqnarray}
G&=&\frac{B(r)h(r)+a^2\cos^2\theta}{\Sigma}.\label{o}
\end{eqnarray}
Thus the rotating black hole solution in Boyer-Lindquist coordinates turns out to be
\begin{eqnarray}\nonumber
ds^2&=&-\frac{B(r)h(r)+a^2\cos^2\theta}{\left(k(r)+a^2\cos^2\theta\right)^2}\Sigma{dt}^2+2a\sin^{2}\theta\frac{B(r)h(r)-k}{\left(k(r)+a^2\cos^{2}\theta\right)^2}\Sigma{dtd\phi}+\frac{\Sigma}{B(r)h(r)+a^2}dr^2+\Sigma{d\theta^2}\\\nonumber
&+&\Sigma\sin^2\theta\left[1+a^2\sin^2\theta\frac{2k(r)-B(r)h(r)+a^2\cos^2\theta}{\left(k(r)+a^2\cos^2\theta\right)^2}\right]d\phi^2.
\end{eqnarray}
Since $A(r)=B(r)$, so $k(r)=h(r)$. Comparison of Eq. (\ref{8}) and Eq. (\ref{o}) gives $\Sigma=r^2+a^2\cos^2\theta$. Hence the rotational solution of black hole in an asymptotically safe gravity theory becomes
\begin{eqnarray*}
(ds)^2&=&-\Big(1-\frac{2Mr}{\Sigma}\Big(1-\frac{\tilde{\xi}}{r^2}\Big)\Big)dt^2-\frac{4aMr\sin^2\theta}{\Sigma}\Big(1-\frac{\tilde{\xi}}{r^2}\Big)dtd\phi+\frac{\Sigma}{\Delta}dr^2+\Sigma{d\theta^2}\\
&+&\sin^2\theta\bigg[r^2+a^2+\frac{2a^2Mr}{\Sigma}\sin^2\theta\Big(1-\frac{\tilde{\xi}}{r^2}\Big)\bigg]d\phi^2,
\end{eqnarray*}
where $\Delta=r^2-2Mr+\frac{2M\tilde{\xi}}{r}+a^2$. This metric reduces to its static and spherically symmetric version when $a\rightarrow{0}$. It is worth to be noted here that the above metric coincides exactly with the one given in section II. This makes our results more reliable.

\end{document}